\documentclass[lettersize,journal]{IEEEtran}
\usepackage{amsmath,amsfonts}
\usepackage{algorithmic}
\usepackage{array}
\usepackage[caption=false,font=normalsize,labelfont=rm,textfont=rm]{subfig}
\usepackage{textcomp}
\usepackage{stfloats}
\usepackage{url}
\usepackage{verbatim}
\usepackage{graphicx}
\def\BibTeX{{\rm B\kern-.05em{\sc i\kern-.025em b}\kern-.08em
		T\kern-.1667em\lower.7ex\hbox{E}\kern-.125emX}}
\usepackage{balance}
\usepackage{algorithm}
\usepackage{color}
\usepackage{multirow}
\usepackage{hyperref} 
\hypersetup{
	colorlinks=true,
	linkcolor=blue,
	citecolor=blue,
	urlcolor=blue
}
\usepackage{mathrsfs}
\usepackage{cite}
\usepackage{booktabs}

\begin{document}

\title{Towards Interpretable PolSAR Image Classification: Polarimetric Scattering Mechanism Informed Concept Bottleneck and Kolmogorov-Arnold Network}

\author{Jinqi Zhang, Fangzhou Han, Di Zhuang, Lamei Zhang,~\IEEEmembership{Senior Member,~IEEE}, Bin Zou,~\IEEEmembership{Senior Member,~IEEE} and Li Yuan

\thanks{This work was supported in part by the National Natural Science Foundation of China (62271172). \textit{(Corresponding author: Di Zhuang; Lamei Zhang)}.  }
\thanks{ J. Zhang, F. Han, L. Zhang, B. Zou and D. Zhuang are with the Department of Information Engineering, Harbin Institute of Technology, Harbin, China.(e-mail: fangzhouhan0018@163.com)}
\thanks{L. Yuan is with the National Key Laboratory of Scattering and Radiation. (ayuanli@163.com) }

\thanks{J. Zhang and F. Han contributed equally to this paper.}
}

\markboth{Journal of \LaTeX\ Class Files,~Vol.~14, No.~8, August~2021}%
{Shell \MakeLowercase{\textit{et al.}}: A Sample Article Using IEEEtran.cls for IEEE Journals}

\maketitle

\begin{abstract}
In recent years, Deep Learning (DL) based methods have received extensive and sufficient attention in the field of PolSAR image classification, which show excellent performance. However, due to the ``black-box" nature of DL methods, the interpretation of the high-dimensional features extracted and the backtracking of the decision-making process based on the features are still unresolved problems. In this study, we first highlight this issue and attempt to achieve the interpretability analysis of DL-based PolSAR image classification technology with the help of Polarimetric Target Decomposition (PTD), a feature extraction method related to the scattering mechanism unique to the PolSAR image processing field. In our work, by constructing the polarimetric conceptual labels and a novel structure named Parallel Concept Bottleneck Networks (PaCBM), the uninterpretable high-dimensional features are transformed into human-comprehensible concepts based on physically verifiable polarimetric scattering mechanisms. Then, the Kolmogorov-Arnold Network (KAN) is used to replace Multi-Layer Perceptron (MLP) for achieving a more concise and understandable mapping process between layers and further enhanced non-linear modeling ability. The experimental results on several PolSAR datasets show that the features could be conceptualization under the premise of achieving satisfactory accuracy through the proposed pipeline, and the analytical function for predicting category labels from conceptual labels can be obtained by combining spline functions, thus promoting the research on the interpretability of the DL-based PolSAR image classification model.

\end{abstract}


\section{Introduction}
\IEEEPARstart{B}{enefiting} from its capability to operate under all-weather, day-and-night conditions, Polarimetric Synthetic Aperture Radar (PolSAR) provides abundant information about the Earth's surface. As an important application, PolSAR image classification is a crucial research area in PolSAR image interpretation and has been extensively applied in geological exploration, terrain surface mapping, vegetation monitoring, urban planning, and ocean surveillance \cite{yingyong1,yingyong2,yingyong3}.

In recent years, Deep Learning (DL) has emerged as the predominant method for PolSAR image classification. Models such as Convolutional Neural Networks (CNNs) \cite{cnn1,cnn2,cnn3}, Recurrent Neural Networks (RNNs) \cite{rnn1,rnn2}, and Transformers \cite{t1,t2} have achieved significant progress. However, many of these approaches rely on complex, high-dimensional features, and their decision-making processes often lack interpretability. As the ``black box" nature of DL, coupled with the current limitations of human cognitive understanding, complicates the interpretation of the semantic meaning of high-dimensional features extracted by these models \cite{int}. Consequently, articulating the rationale behind feature-driven decisions of DL models remains challenging. This inherent opacity raises significant concerns regarding the reliability, security, and stability of DL applications in real-world scenarios \cite{inter1,inter2,inter3}. Similar issues have arisen in PolSAR image classification. Most existing DL models for PolSAR analysis follow a common paradigm: a feature extraction network processes the input PolSAR data to generate deep features, which are subsequently mapped to class labels via a Multi-Layer Perceptron (MLP). Despite their impressive performance, understanding how these models extract and utilize information from PolSAR data—and clarifying the principles underlying their transformation of high-dimensional features into final classifications—remains an unresolved challenge.

Luckily, unlike optical images, which represent targets through RGB channel values, PolSAR data are generated by the active imaging mechanism of the polarimetric SAR system. This unique data structure captures detailed information about the interaction between electromagnetic waves and observed targets. Consequently, PolSAR data can be analyzed using specialized Polarimetric Target Decomposition (PTD) methods \cite{PTd1,PTd2,PTd3,PTd4}, which relate target structure to physical scattering mechanisms. For example, the widely used Freeman–Durden three-component decomposition \cite{Freeman} factorizes the coherency matrix of PolSAR data into three primary components: surface scattering, double-bounce scattering and volume scattering. Surface scattering typically originates from flat surfaces such as bare ground. Double-bounce scattering arises from structures like building walls and ground planes or utility poles. Volume scattering results from volumetric scatterers such as vegetation canopies. The decomposition provides a physically interpretable framework for analyzing PolSAR data based on scattering characteristics. 

As a representation of physical scattering mechanisms, PTD methods have been instrumental in applications such as PolSAR data augmentation, self-supervised learning, and model development that integrates physical priors \cite{self,self1}. Multiple researches support their effectiveness in enhancing classification performance. However, while many studies emphasize the performance benefits of incorporating PTD methods, they often overlook their intrinsic interpretability and value as a human-understandable representation with clear physical meaning. Based on these insights, we pose the following research question:

\textit{Could PTD enhance the interpretability of DL-based PolSAR image classification models by refining high-dimensional semantic features and aligning model decision-making with human understanding?}

To address this challenge, our work aims to enhance the interpretability of PolSAR image classification models by integrating polarimetric scattering information. Specifically, we propose the use of a Concept Bottleneck Model (CBM) framework \cite{Koh}. CBMs \cite{cbm_Editable,cbm_post-hoc} are among the most prominent interpretable DL approaches, consisting of two main components: an intermediate Concept Bottleneck Layer (CBL), where neurons correspond to human-interpretable concepts, and a subsequent linear decision layer that combines these concepts for final classification. In the context of PolSAR classification, we convert well-established PTD outputs into concept labels, thereby shifting from the traditional ``feature-label" paradigm to a ``feature-concept-label" framework that better aligns with human cognitive reasoning. Furthermore, to enhance the model’s capacity to capture nonlinear dependencies and provide transparent inference, we replace the standard MLP with the Kolmogorov–Arnold Network (KAN) \cite{2024KAN}. This architecture improves both model performance and interpretability. The primary contributions of this paper are summarized as follows:

\begin{enumerate}
    \item We firstly focus on the interpretability of high-dimensional features and decision-making processes in DL-based PolSAR image classification models, proposing an effective framework that utilizes the ability of PTD to interpret physical mechanisms.
    \item We propose an effective method for constructing conceptual labels for the CBM interpretability framework, based on the polarimetric scattering mechanism as the foundation.
    \item We replace the MLP in the common inference chain with KAN, which is more conducive to capturing nonlinearity and enhancing the interpretability of the inference model.
\end{enumerate}

\section{Related Works}

\subsection{PolSAR Image Classification Method}

Due to the abundant phase information of PolSAR data compared to single-channel SAR data, PolSAR image classification has garnered considerable attention from researchers. Hänsch \textit{et al.} \cite{Hänsch} employed the random forest algorithm to classify PolSAR images, investigating the effects of feature selection and classification within this framework.However, traditional classification methods often fail to effectively capture the complex scattering mechanisms inherent in PolSAR data. Zhou \textit{et al.} \cite{Zhou} first converted multi-view PolSAR data in the form of coherent or covariance matrices into normalized six-dimensional real-valued feature vectors, which were then fed into a four-layer CNN specifically designed for PolSAR classification. To more effectively extract features and integrate both polarimetric and spatial domain characteristics, Yang \textit{et al.} \cite{Yang} proposed a method that combines ResNet with a deep autoencoder. Dong \textit{et al.} \cite{Dong} introduced a Vision Transformer (ViT)-based representation learning framework incorporating both supervised and unsupervised learning, achieving performance superior to that of CNNs. Alkhatib \textit{et al.} \cite{Alkhatib} developed Hybrid CVNet, a novel architecture that combined CNN and ViT to extract complementary information and efficiently capture interdependencies within the data. Kuang \textit{et al.} \cite{Kuang} proposed a polarimetry-inspired contrastive learning approach that integrated unsupervised complex-valued contrastive learning with a hybrid anti-imbalance strategy. Zhang \textit{et al.} \cite{zhang} introduced MCDiff, a multilevel conditional diffusion model that incorporated prior scattering knowledge, multiscale noise prediction, and high-order statistical feature learning. Li \textit{et al.} \cite{Li} developed a hybrid complex-valued 2D/3D CNN with an attention mechanism to enhance both feature extraction and computational efficiency. Shi \textit{et al.} \cite{Shi} proposed LRCM, a lightweight model that learnd covariance matrix features directly in Riemannian space. Wang \textit{et al.} \cite{Wang} designed a multi granularity hybrid CNN-ViT model that fuses local feature extraction from CNNs with global multiscale attention from ViTs. Das \textit{et al.} \cite{Das} presented a two-step PolSAR classification method that integrated multiscale SVD profiles with a lightweight dual-branch CNN.

Despite these advancements, most existing DL-based methods prioritize performance optimization while largely overlooking model interpretability. In this work, we aim to improve the interpretability of DL-based PolSAR image classification models without compromising accuracy, striving to achieve a balance between predictive performance and explainability.

\subsection{PTD Methods and Applications}
As an effective and widely used strategy for interpreting PolSAR imagery, PTD enables the extraction of distinctive polarimetric characteristics from complex scattering signals. These characteristics reflect the physical properties of targets and significantly enhance classification performance.

 In the realm of coherent target decomposition, Krogager \textit{et al.} \cite{Krogager} proposed a decomposition method that broke the polarization scattering matrix into the sum of three coherent components with clear physical significance, based on the circular polarization basis. Cameron \textit{et al.} \cite{Cameron} introduced a decomposition approach grounded in the symmetry and reciprocity of scattering objects, enabling the identification of typical target types. In the realm of incoherent target decomposition, Freeman and Durden \textit{et al.} \cite{Freeman} initially proposed such methods based on the covariance matrix \( C \) or coherence matrix \( T \). Yamaguchi \textit{et al.} \cite{Yamaguchi} further extended this framework by incorporating a fourth component—spiral scattering—to better characterize man-made structures, thereby improving applicability in complex environments such as urban areas.

PTD results typically capture the physical properties of objects, making them widely used to improve classification accuracy. Zhou \textit{et al.} \cite{ZhouSVM} combined Support Vector Machines (SVM) with \( H \) and \( \alpha \) parameters to enhance target classification. Imani \textit{et al.} \cite{imani} proposed PSI, an iterative method that refined SVM labels by incorporating spatial and polarimetric similarities, leading to improved PolSAR classification. Qu \textit{et al.} \cite{tian1} introduced an unsupervised classification method that fused high-resolution single-pol and medium-resolution PolSAR data. By leveraging Freeman–Durden and H/Alpha-Wishart decomposition-based methods, their approach highlights the complementary nature of physical scattering characteristics and image features. Hua \textit{et al.} \cite{Hua1} proposed a deep fusion network that integrated PolSAR scattering characteristics with physical features to enhance classification accuracy and interpretability, even with limited labeled data. Zhang \textit{et al.} \cite{zz1} developed a contrastive learning method that utilized polarimetric information to improve PolSAR classification under data-scarce conditions.

However, existing work primarily focuses on using PTD as a feature extraction method to enhance tasks such as classification and detection, often overlooking its potential as an effective tool for improving interpretability.

\subsection{Interpretable Deep Learning Models}

Due to the opaque nature of neural networks, extensive research has been devoted to interpretable neural networks, which are generally classified into two main categories: post hoc and ante hoc approaches.

Among post hoc methods, Zhou \textit{et al.} \cite{ZhouLearningdeep} proposed Class Activation Mapping (CAM), which visually interprets CNNs using attribution heatmaps. An extension of CAM, known as Gradient-weighted Class Activation Mapping (Grad-CAM), was introduced by Selvaraju \textit{et al.} \cite{Grad-cam}. Post hoc interpretability primarily functions as a corrective tool for pre-trained, complex models. However, it has limitations, including imprecise, unreliable, or unintuitive interpretations, particularly in applications requiring transparency and real-time trust. By contrast, ante hoc approaches aim to enhance interpretability during model training. Koh \textit{et al.} \cite{Koh} introduced the CBM, which predicts target tasks via a conceptual layer. Numerous models have since been developed to improve upon the foundational principles of CBM. To address the inherent trade-off between interpretability and accuracy in CBMs, Espinosa \textit{et al.} \cite{CEM} proposed the Concept Embedding Model (CEM), which provides semantically meaningful concept representations and effective intervention capabilities while maintaining high accuracy, particularly under limited concept supervision. Because CBMs typically require large amounts of manually annotated concept labels, Oikarinen \textit{et al.} \cite{Oikarinen} introduced a label-free CBM that automates concept annotation using the CLIP model. CBMs have also been applied in various domains: Cristiano \textit{et al.} \cite{skin} utilized CBMs for skin lesion diagnosis, while Wong \textit{et al.} \cite{Wong} explored their use for inherent decision interpretability in DL-based automatic modulation classification. To the best of our knowledge, no prior study has applied CBMs to PolSAR classification.

Recent researches have also focused on the interpretability of models applied on SAR images. Datcu \textit{et al.} \cite{tian4} advocated for the integration of explainable artificial intelligence into SAR data interpretation by leveraging established physical models for interpretable data transformation. Huang \textit{et al.} \cite{tian3} introduced PGIL, a physics-guided DL framework for SAR image classification that enhances both interpretability and performance under limited labeled data by injecting physics-aware features. Zhang \textit{et al.} \cite{dual} proposed DAP-Net, a dual-channel SAR target recognition network that integrates attention mechanisms with polarimetric information from dual polarizations, thereby improving feature extraction and interpretability. Ge \textit{et al.} \cite{kejieshi1} developed a SHAP value-guided explanation model for PolSAR data, combining a physically interpretable feature module with a spatial interpretation network.

However, although the aforementioned methods have improved the interpretability of SAR and PolSAR images to some extent, they still encounter challenges such as limited intuitive comprehensibility for human users and fixed model structure. In contrast, the approach proposed in this paper integrates PTD with the CBM, leveraging explicit model-irrelevant physical concept bottlenecks to achieve both physical interpretability and model flexibility.

\section{Methodology}

\subsection{Polarimetric Target Decomposition}
The concept label is the most critical component of the CBM. To extract characteristic attribute values for typical targets, we employ the Cloude–Pottier decomposition \cite{tezhengzhi}, Freeman–Durden decomposition \cite{Freeman}, and Huynen decomposition \cite{huynen}.

Based on the properties of PolSAR data, each pixel in a PolSAR image can be represented by a complex scattering matrix as follows:
\begin{equation}
    \begin{split}
      & S = \begin{bmatrix}
S_{HH} & S_{HV} \\
S_{VH} & S_{VV}
\end{bmatrix} \\
    \end{split}
    \label{matrix}
\end{equation}

And the Polarization Power can be expressed as:
\begin{equation}
    \begin{split}
      & S_{\text{span}} = |S_{HH}|^2 + 2 |S_{HV}|^2 + |S_{VV}|^2 \\
    \end{split}
    \label{span}
\end{equation}

where \( S_{PQ}(P, Q \in \{H, V\}) \) represents the backscattering coefficient of the polarized electromagnetic wave in the direction of the \( Q \) and the direction of the receiving \( P \). \( H \) and \( V \) represent the horizontal and vertical polarizations, respectively.

To obtain the coherence matrix from the scattering matrix, the latter must first be vectorized. This vectorization is commonly performed using the Pauli basis vectors. The Pauli basis vector is defined as follows \cite{Puail}:
\begin{equation}
    \begin{split}
      & \vec{k} = \frac{1}{\sqrt{2}} [S_{HH} + S_{VV}, S_{HH} - S_{VV}, 2S_{HV}]' \\
    \end{split}
    \label{vector_k}
\end{equation}

The polarization coherence matrix(\( T \)) is obtained by the outer product of Pauli's basis vectors, which is defined as:
\begin{equation}
    \begin{split}
      & T = \langle \vec{k} \vec{k}^H \rangle
    \end{split}
    \label{matrix_T}
\end{equation}

Noting that \( T \) is a Hermitian matrix, all its elements are complex numbers, with the diagonal elements being real. The decomposition methods described below rely on this property.

\textit{1) Cloude-Pottier Decomposition:} Cloude-Pottier Decomposition extracts polarization information by calculating the eigenvalues and eigenvectors of the coherence matrix and then calculate parameters such as Entropy (\( H \)), Anisotropy (\( A \)), and Alpha Angle (\( \alpha \)). The coherence matrix \( T \) could be expressed through eigenvalue decomposition \cite{tezhengzhi} as follows:
\begin{equation}
\textbf{T} = \textbf{U} \boldsymbol{\Lambda} \textbf{U}_H.
\end{equation}

Where $\boldsymbol{\Lambda}$ is a diagonal matrix containing the eigenvalues $\lambda_1, \lambda_2, \lambda_3$, and $\lambda_1 \geq \lambda_2 \geq \lambda_3$; $\textbf{U}$ is the eigenvector matrix. Cloude-Pottier decomposition characterizes target scattering using these three parameters: \( H\), \( \alpha \) and \( A \).
\begin{align}
H &= -\sum_{i=1}^{3} P_n \log_3 P_n \quad
\bar{\alpha} = \sum_{n=1}^{3} P_n \alpha_n \quad
A = \frac{\lambda_2 - \lambda_3}{\lambda_2+\lambda_3}
\end{align}

where \(P_n = \lambda_n / (\lambda_1 + \lambda_2 + \lambda_3)\) and \( \alpha \) is related to the matrix \( U \). The parameter \( H \) quantifies the complexity of a target's scattering mechanism, ranging from 0 (single scattering, e.g., water) to 1 (complex scattering, e.g., urban areas). The parameter \( \alpha \) represents the average scattering type: values near $0^{\circ}$ indicate surface scattering, around $45^{\circ}$ suggest double-bounce, and near $90^{\circ}$ imply volume scattering. The anisotropy parameter \( A \) helps distinguish scattering types in complex, high-entropy regions. 

\textit{2) Freeman-Durden Decomposition:} The polarization covariance matrix of the target is modeled as the sum of three scattering mechanisms: volume scattering, double-bounce scattering, and surface scattering. 

Freeman–Durden decomposition divides the SPAN into the sum of three typical scattering mechanisms:
\begin{align}
P_S &= f_S(1 + |\beta|^2), \quad
P_D = f_D(1 + |\alpha|^2), \quad
P_v = \frac{8f_v}{3} 
\end{align}

where \( f\) represents the corresponding scattering coefficient, \( \alpha \) and \( \beta\) is related to polarization covariance matrix. The three scattering components from the Freeman decomposition serve as conceptual labels: high surface scattering corresponds to smooth areas; strong double-bounce indicates vertical structures such as buildings or trees; and dominant volume scattering reflects the internal features of vegetation, forests, or other random media.

\textit{3) Huynen Decomposition:} The fundamental principle of the Huynen decomposition method is to separate the echo data into two components: one representing the identified target and the other representing the participating or noise component. During the decomposition process, nine parameters can be derived, each containing specific target scattering information. According to the Huynen decomposition, the coherence matrix \( T \) is expressed in the following form:
\begin{equation}
    \begin{split}
        T = \begin{bmatrix}
            2A_0 & C - jD & H + jG \\
            C + jD & B_0 + B & E + jF \\
            H - jG & E - jF & B_0 - B
        \end{bmatrix}
    \end{split}
    \label{matrix}
\end{equation}

The variables \( A_0, B_0, B, C, D, E, F, G \) and \( H \) are collectively referred to as Huynen parameters. These nine target parameters are not constrained by specific model assumptions and each encapsulates critical physical characteristics of the observed target, making them essential for comprehensive analysis. For example, \( A_0 \) and \( B_0 \) represent the total scattered power from the target's smooth and irregular rough components, respectively. Analyzing the power ratio between these two parameters provides insights into the target's symmetry and degree of structural regularity.

\subsection{Polarimetric Concept Label Construction}

\begin{figure*}[t]
    \centering
    \includegraphics[width=1\linewidth]{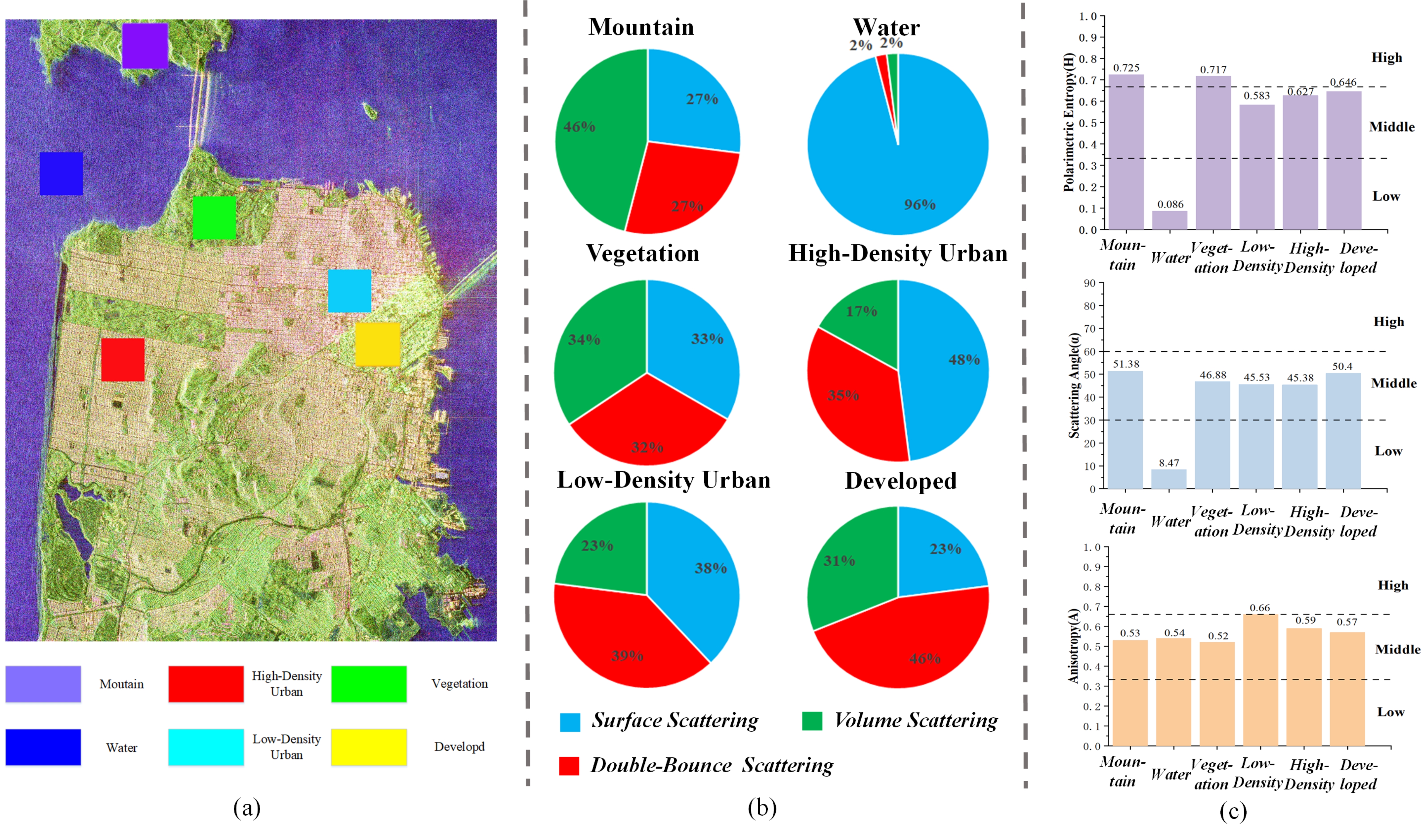}
    \caption{The conceptual labels constructed by the PTD  of six typical terrain surface categories in GF-3 dataset. (a) Pixel block selection for PTD (b) Cloude-Pottier decomposition results: result of polarimetric entropy, scattering angle and anisotropy, respectively. (c) Freeman-Durden decomposition results, the proportion results of the three polarization components for six typical land cover types.}
    \label{PTD result}
\end{figure*}

In the previous section, we introduced the PTD methods and derived relevant polarimetric features. Focusing on specific terrain surface categories, we applied PTD to obtain corresponding concept descriptions and subsequently constructed polarimetric concept labels. Although the datasets used in the experiments were acquired from different frequency bands—namely C-band and L-band—our study emphasizes the band-robust polarization characteristics of typical features, that is, the features chosen and the the corresponding polarization scattering mechanism are independent of the sensor band. We conducted an experimental analysis on GF-3 data for an example, covering mountain, water, vegetation, high-density urban, low-density urban, and developed areas \cite{polSF}. As a result, we summarized concept labels based on \(S_{\text{span}}\), degree of polarization, polarization mode, Cloude–Pottier decomposition, Freeman-Durden decomposition, and Huynen decomposition.

To ensure the conciseness of the PTD decomposition and to prevent poor-quality samples from affecting the results, we manually selected 200×200 pixel patches representing six typical land cover types, as illustrated in Fig. \ref{PTD result}(a). The concept labels in this study were constructed based on domain knowledge and statistical analysis. Label construction for these parameters generally followed two approaches: analyzing relative proportions within the same land cover type and examining absolute parameter values across different land cover types. We illustrated the concept label construction process using the Freeman-Durden and Cloude-Pottier decompositions as examples. The results for six representative land cover types within the selected image patches are shown in Fig. \ref{PTD result}(b) and Fig. \ref{PTD result}(c), respectively.

For Freeman-Durden decomposition, existing studies\cite{yuzhi} suggest that analyzing polarization proportions should not rely solely on absolute values but also consider the relative magnitudes between different scattering components. Accordingly, we define two rules as follows:
\begin{equation}
    \begin{split}
      & P_{max} \geq 0.4 \quad\land \quad P_{max} - P_{2nd} \geq 0.1
    \end{split}
    \label{yuzhi1}
\end{equation}
\begin{equation}
    \begin{split}
      & P_{i} \geq 0.5 P_{max} 
    \end{split}
    \label{yuzhi2}
\end{equation}

Where \(P_{i}\) denotes the proportion of the \(i\)-th scattering component, \(P_{max}\) indicates the component with the highest scattering power, and \(P_{2nd}\) refers to the component with the second-highest power. The dominant scattering component is determined by Equation \ref{yuzhi1}: if the condition is satisfied, the component with the highest power proportion is labeled as dominant; otherwise, it is classified as secondary scattering. For the other two scattering mechanisms, if they satisfy Equation \ref{yuzhi2}, they are classified as secondary scattering; if not, they are defined as weak scattering. Taking High-Density Urban as an example, surface scattering meets the criterion in Equation \ref{yuzhi1} and is labeled as ``surface scattering dominant," double-bounce scattering satisfies Equation \ref{yuzhi2} and is labeled as “secondary double-bounce scattering,” while volume scattering is labeled as ``weak volume scattering."

For concept label construction in Cloude-Pottier decomposition, the parameters 
\(H\) (ranging from 0 to 1), \( \alpha \)(ranging from $0^{\circ}$ to $90^{\circ}$), and \(A\)(ranging from 0 to 1) are each divided into three equal intervals and labeled as ``High," ``Middle," and ``Low," respectively. For example, the mountain corresponds to high polarization entropy, medium scattering angle, and medium \(A\). The construction of other derived parameters follows a similar approach. Table \ref{concept labels} summarizes all constructed concept labels.
\begin{table*}[t]
\centering
\caption{Concept labels of typical terrain surface polarimetric features obtained through PTD}
\begin{tabular}{c>{\raggedright\arraybackslash}p{\dimexpr\textwidth-3cm\relax}}
\specialrule{1.2pt}{0pt}{0pt}
\rule{0pt}{13pt}\textbf{\small Category} & \textbf{\normalsize Concept Labels} \\
\midrule
 \textbf{\footnotesize Mountain}& Volume scattering dominant, Secondary surface scattering, Secondary double-bounce scattering, Cross polarization dominant, Low degree of polarization, High polarization entropy, Medium anisotropy, Medium scattering angle, Irregular target, Low scattering power, Asymmetric \\
\midrule
 \textbf{\footnotesize Water} & Surface scattering dominant, Weak volume scattering, Weak double-bounce scattering, Vertical polarization dominant, High degree of polarization, Low polarization entropy, Medium anisotropy, Small scattering angle, Highly regular target, Medium scattering power, High symmetry \\
\midrule
 \textbf{\footnotesize Vegetation} & Secondary surface scattering, Secondary volume scattering, Secondary double-bounce scattering, Horizontal polarization dominant, Medium degree of polarization, High polarization entropy, Medium anisotropy, Medium scattering angle, Regular and irregular parts equivalent, Low scattering power, Medium symmetry \\
\midrule
 \textbf{\footnotesize High-Density Urban} & Surface scattering dominant, Secondary double-bounce scattering, Weak volume scattering, Horizontal polarization dominant, Medium degree of polarization, Medium polarization entropy, Medium anisotropy, Medium scattering angle, Regular and irregular parts equivalent, Medium scattering power, Medium symmetry \\
\midrule
 \textbf{\footnotesize Low-Density Urban} & Secondary double-bounce scattering, Secondary surface scattering, Secondary volume scattering, Horizontal polarization dominant, High degree of polarization, Medium polarization entropy, High anisotropy, Medium scattering angle, Regular and irregular parts equivalent, High scattering power, Medium symmetry \\
\midrule
 \textbf{\footnotesize Developed} & Double-bounce scattering dominant, Secondary volume scattering, Secondary surface scattering, Vertical polarization dominant, Medium degree of polarization, Medium polarization entropy, Medium anisotropy, Medium scattering angle, Irregular target, High scattering power, Asymmetric \\
\specialrule{1.2pt}{0pt}{0pt}
\end{tabular}
\label{concept labels}
\end{table*}

Through PTD, we constructed concept labels that were intrinsically linked to the physical scattering characteristics of targets. These labels not only capture the underlying physical properties of the observed objects but also facilitate clear differentiation among various land cover types.

\subsection{Concept Bottleneck Model with Kolmogorov-Arnold Network}
By integrating the PTD mechanism unique to PolSAR data, we have constructed polarimetric concept labels and developed an interpretable PolSAR image classification model. This model addresses two critical challenges often overlooked by conventional DL-based classifiers: the interpretability of high-dimensional features and the traceability of decision-making processes. These objectives are achieved through the effective integration of two interpretability frameworks—CBM and KAN—within the classification architecture.

The study start with the structure and computing approach of the base CBM. Considering a dataset comprising $N$ pairs of image samples and labels, which could be denoted as: $\mathscr{D}=\{{{X}_{n}},{{y}_{n}}\}_{n=1}^{N}$, where the ${{X}_{n}}\in {{\mathbb{R}}^{h\times w\times c}}$ and ${{y}_{n}}\in {{\{0,1\}}^{C}}$. A basic classification model that includes feature extraction and MLP layer mapping can be abstracted as follows:
\begin{equation}
    \begin{split}
      & {{{\hat{y}}}_{n}}=\sigma (W\cdot \mathcal{F}({{X}_{n}})+b) \\ 
 & \mathscr{L}_{cls}=-\sum\limits_{n=1}^{N}{{{y}_{n}}\cdot \log ({{{\hat{y}}}_{n}})} \\ 
    \end{split}
    \label{main}
\end{equation}

Where the function $\mathcal{F}(\cdot)$ refers to the feature extraction model. The $\sigma(\cdot)$, $W$ and $b$ refers to the activation function, learnable Weight and bias term of MLP layer. $\mathscr{L}$ is the cross-entropy loss between the predictions and the labels. On this basis, CBM adds CBL between the feature and the predictions. Considering concept set $\mathscr{C}=\{{{c}_{1}},{{c}_{2}},\cdots {{c}_{K}}\}$, comprising $K$ concepts. Using 0/1 to represent the correspondence between different categories and concept sets, the original dataset $\mathscr{D}$ can be expanded to ${{D}^{c}}=\{{{X}_{n}},{{y}_{n}},{{\mathbf{c}}_{n}}\}$, where the ${{\mathbf{c}}_{n}} \in {{\mathbb{R}}^{K}}$ is the vector of the concept labels. Unlike the paradigm of directly mapping features to the labels, CBM first maps features to the concept label layer and maps category labels based on predictions from concept labels, which could be denote as follows:
\begin{equation}
    {{\hat{y}}_{n}}={{g}_{c\to t}}({{g}_{f\to c}}(\mathcal{F}({{X}_{n}})))
\end{equation}

Where, the functions ${{g}_{f\to c}}:{{\mathbb{R}}^{D}}\to {{\mathbb{R}}^{K}}$ and ${{g}_{c\to t}}:{{\mathbb{R}}^{D}}\to {{\mathbb{R}}}$ refer to the process of mapping features to the concept and mapping concept to the category logits, respectively. As can be seen from the above calculation process, since the concept has been processed into the form of one-hot labels, the overall optimization goal is not only the cross entropy loss between category logits and labels, but also the optimization goal for concept labels. For the latter, we use the typical BCEWithLogitsLoss, which combines the Binary Cross Entropy (BCE) loss and a sigmoid layer:
\begin{equation}
    \begin{split}
        \mathscr{L}_{con} &= \frac{1}{N\times K}\sum_{j = 1}^{N}\sum_{i = 1}^{K} \left[ c_{ij}\log\left(1 + e^{-\hat{c}_{ij}}\right) \right. \\
                          & \quad \left. + (1 - c_{ij})\log\left(1 + e^{\hat{c}_{ij}}\right) \right]
    \end{split}
\end{equation}

With the introduction of the CBL and the refinement of concept labels, human-interpretable concepts can be effectively leveraged to analyze feature-relevant information. When a concept plays a significant role in the model’s classification decision, its predicted value tends to more closely align with the ground truth. However, although current CBMs can qualitatively indicate the importance of each concept in classification, they still fall short in explaining the underlying decision-making process—specifically, how the model integrates concept label information to generate the final prediction. This lack of transparency primarily stems from the MLP layer typically used in CBMs.

Due to the above reasons, our work uses KAN to replace MLP, thus overcoming the destructive effect of MLP on the interpretability of CBM, in order to achieve the purpose of traceability model decision-making process. KAN is grounded in the Kolmogorov-Arnold Representation Theorem (KART) proposed by the Russian mathematician Andrey Kolmogorov in 1957, which asserts that any continuous multivariable function can be expressed as a finite superposition of continuous univariate functions $f({{x}_{1}},{{x}_{2}},{{x}_{3}},\cdots ,{{x}_{n}})$. More specifically, for a smooth $f:{{[0,1]}^{n}}\to \mathbb{R}$:
\begin{equation}
 f(x)=f({{x}_{1}},{{x}_{2}},{{x}_{3}},\cdots ,{{x}_{n}})=\sum\limits_{q=1}^{2n+1}{{{\Phi }_{q}}(\sum\limits_{p=1}^{n}{{{\phi }_{q,p}}({{x}_{p}})})}
\end{equation}

Where ${{\phi }_{q,p}}:[0,1]\to \mathbb{R}$ and ${{\Phi }_{q}}:\mathbb{R}\to \mathbb{R}$ are continuous univariate functions that map input variables into a result through a sum of simpler functions. 

Specifically, the KAN layer can be defined as a 1D function matrix by unifying the outer functions ${{\Phi }{q}}$ and the inner functions ${{\phi }{q,p}}$ into a series of KAN layers with $n_{\text{in}}$ input nodes and $n_{\text{out}}$ output nodes, as follows:
\begin{equation}
\Phi =\{{{\phi }_{q,p}}\},\text{  }p=1,2,3,\cdots {{n}_{in}},\text{  }q=1,2,3,\cdots {{n}_{out}}
\end{equation}

Suppose a KAN with $L$ layers, the number of nodes in each layer can be expressed as $[{{n}_{0}},{{n}_{1}},\cdots ,{{n}_{L-1}}]$, the activation value of the $(l+1,j)$ neuron is simply the sum of all incoming post-activations:
\begin{equation}
{{x}_{l+1,j}}=\sum\limits_{i=1}^{{{n}_{l}}}{{{\phi }_{l,j,i}}({{x}_{l,j}})},\text{  }j=1,2,\cdots ,{{n}_{l+1}}
\end{equation}

\begin{figure*}[t]
    \centering
    \includegraphics[width=1\linewidth]{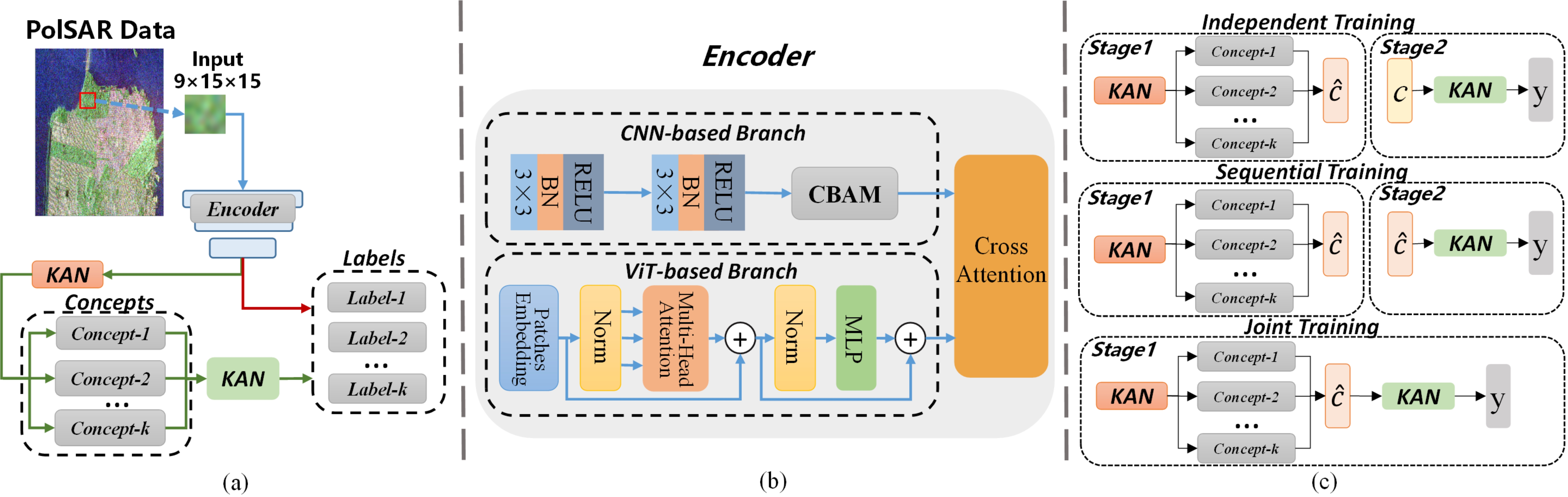}
    \caption{The overall model of the proposed method. (a) the proposed PaCBM. The red data stream represents the calculation of category labels directly from image features, and the green data stream represents the process of mapping concept labels from image features, and then predicting category labels through the predicted value of concept labels. (b) The encoder architecture used in this work, which adopts a dual-branch structure combining CNN-based and ViT-based networks with a cross-attention mechanism. (c) the three strategies for training concept label mapping in PaCBM. }
    \label{fig:model}
\end{figure*}

In matrix form, it could be expressed as follows:
\begin{equation}
{{x}_{l+1}}=\left( \begin{matrix}
   {{\phi }_{l,1,1}}(\cdot ) & {{\phi }_{l,1,2}}(\cdot ) & \cdots  & {{\phi }_{l,1,{{n}_{l}}}}(\cdot )  \\
   {{\phi }_{l,2,1}}(\cdot ) & {{\phi }_{l,2,2}}(\cdot ) & \cdots  & {{\phi }_{l,2,{{n}_{l}}}}(\cdot )  \\
   \vdots  & \vdots  & \cdots  & \vdots   \\
   {{\phi }_{l,{{n}_{l+1}},1}}(\cdot ) & {{\phi }_{l,{{n}_{l+1}},2}}(\cdot ) & \cdots  & {{\phi }_{l,{{n}_{l+1}},{{n}_{l}}}}(\cdot )  \\
\end{matrix} \right){{x}_{l}}
\end{equation}

Unified input and output representation makes it possible to stack KAN layers to construct deeper and wider KAN. Therefore, for an $L$-layer KAN, suppose the input vector be $x\in {{\mathbb{R}}^{{{n}_{0}}}}$, the calculation process could be denoted as follows:
\begin{equation}
    \mathcal{K}(x)=({{\Phi }_{L-1}}\circ {{\Phi }_{L-2}}\circ \cdots \circ {{\Phi }_{1}}\circ {{\Phi }_{0}})x.
\end{equation}

Compared to the MLP, the KAN simplifies complex relationships between nodes by representing them as a summation of parameterized 1-D functions. This architectural design enhances the model's systematic nature and interpretability, as inter-layer dependencies are modeled through B-spline parameters instead of independently learned weights and biases. In the CBM framework, we replace the MLP layer with KAN to leverage its improved interpretability. Specifically, we employ two separate KANs to model the following two functions: ${{\phi }_{q,p}}:[0,1]\to \mathbb{R}$ and ${{\Phi }_{q}}:\mathbb{R}\to \mathbb{R}$. Thus, in the KAN-CBM, the prediction of concept labels $\hat{c_n}$ and the prediction of category labels $\hat{y_n}$  can be written as follows:
\begin{equation}
\begin{split}
   {{\hat{c}}_{n}} & =\mathcal{K}{_{f\to c}}(\mathcal{F}({{X}_{n}})) \\
{{\hat{y}}_{n}} & =\mathcal{K}_{c\to t}(\mathcal{K}_{f\to c}(\mathcal{F}({{X}_{n}}))) 
\end{split}
\end{equation}

By effectively combining above-mentioned CBM and KAN in the PolSAR interpretability classification model, we can theoretically achieve two basic tasks: human understanding of high-dimensional features in PolSAR image classification and backtracking of decision-making processes.

\subsection{Interpretable PolSAR Image Classification Model}
Although CBMs offer advantages in feature-level interpretability, they often underperform in classification accuracy compared to models optimized solely for predictive performance. In some cases, the use of CBMs may even lead to a substantial decrease in accuracy. We attribute this performance degradation to a key limitation: the reliance on a limited set of predefined concepts results in information loss. Specifically, when high-dimensional features are passed through the CBL, only human-interpretable information is retained, while other potentially critical features—those not easily mapped to predefined concepts—are discarded, reducing the model’s ability to make accurate predictions.

To improve interpretability without significantly sacrificing classification performance, we propose a novel architecture that incorporates CBM as an auxiliary task, serving as a regularization component. This architecture, referred to as Parallel CBM (PaCBM), jointly optimizes both the main and auxiliary tasks by sharing a common feature representation. As shown in Fig. \ref{fig:model}(a), this design balances interpretability and accuracy while mitigating information loss. The classification loss function is denoted as $\mathscr{L}_{main}$, and the concept loss function as $\mathscr{L}_{aux}$. A hyperparameter $\lambda$ is used to balance the contributions of these two losses during model optimization, resulting in the following overall loss function for PaCBM:
\begin{equation}
    \mathscr{L}_{total} = \mathscr{L}_{main} + \lambda\cdot \mathscr{L}_{aux}
\end{equation}

PaCBM effectively mitigates the performance degradation caused by the loss of high-dimensional semantic information. However, to maintain competitive accuracy with state-of-the-art classifiers, a powerful feature encoder remains essential. Inspired by HyCVNet\cite{HybridCVNet}, we design a dual-branch architecture that integrates a CNN and a Vision Transformer (ViT), enhanced with a cross-attention mechanism to efficiently fuse local and global features. The CNN branch comprises two convolutional layers and a Convolutional Block Attention Module (CBAM)\cite{cbam}, as shown in Fig. \ref{fig:model}(b).

Since CBM includes two inference paths—namely, feature-to-concept prediction and concept-to-label classification—it is important to explore their interaction during model training. Specifically, our PaCBM framework supports three distinct training strategies, as illustrated in Fig. \ref{fig:model}(c):

\textit{1)Independent training:}
The independent bottleneck train the $\mathcal{K}_{f\to c}(\cdot)$ and $\mathcal{K}_{c\to t}(\cdot)$ independently:
\begin{equation}
\begin{split}
    {{\hat{\mathcal{K}}}_{f\to c}} &=\arg {{\min }_{{{\mathcal{K}}_{f,c}}}}\sum\nolimits_{n}{{{L}_{C}}(\mathcal{F}({{X}_{n}});{{c}_{n}})} \\
    {{\widehat{\mathcal{K}}}_{c\to t}}&=\arg {{\min }_{{{\mathcal{K}}_{c,t}}}}\sum\nolimits_{n}{{{L}_{Y}}({{\mathcal{K}}_{c\to t}}(c_n);{{y}_{n}})}. 
\end{split}
\end{equation}

During training, $\mathcal{K}_{f\to c}(\cdot)$ and $\mathcal{K}_{c\to t}(\cdot)$ are optimized independently. Stage 1 learns to predict concept labels from input features, while Stage 2 maps ground-truth concept labels to target categories. During inference, the predicted concept labels generated in Stage 1 are used as inputs to Stage 2 to produce the final classification results.

\textit{2)Sequential training:}
The sequential bottleneck train the $\mathcal{K}_{f\to c}(\cdot)$ and $\mathcal{K}_{c\to t}(\cdot)$ in order: 
\begin{equation}
\begin{split}
    {{\hat{\mathcal{K}}}_{f\to c}} &=\arg {{\min }_{{{\mathcal{K}}_{f,c}}}}\sum\nolimits_{n}{{{L}_{C}}(\mathcal{F}({{X}_{n}});{{c}_{n}})} \\
    {{\widehat{\mathcal{K}}}_{c\to t}}&=\arg {{\min }_{{{\mathcal{K}}_{c,t}}}}\sum\nolimits_{n}{{{L}_{Y}}({{\mathcal{K}}_{c\to t}}(\mathcal{F}({{X}_{n}}));{{y}_{n}})}. 
\end{split}
\end{equation}

During training, $\mathcal{K}_{f\to c}(\cdot)$ is trained in the same manner as in independent training. The key difference lies in Stage 2, where, instead of using ground-truth concept labels, the model receives as input the predicted concept labels generated by the Stage 1.

\textit{3)Joint training:}
Joint training means that two functions  $\mathcal{K}_{f\to c}(\cdot)$ and $\mathcal{K}_{c\to t}(\cdot)$ need to be optimized simultaneously:

\begin{equation}
\begin{split}
   \left\langle {{\widehat{\mathcal{K}}}_{f\to c}},{{\widehat{\mathcal{K}}}_{c\to t}} \right\rangle &= \underset{_{{{\mathcal{K}}_{f,c}},{{\mathcal{K}}_{c,t}}}}{\mathop{\arg \min }}\sum\nolimits_{n}{[{{L}_{Y}}({{\mathcal{K}}_{c,t}}({{\mathcal{K}}_{f\to c}}(\mathcal{F}({{X}_{n}})));{{y}_{n}})} \\ 
 & +\lambda {{L}_{C}}({{\mathcal{K}}_{f\to c}}(\mathcal{F}({{X}_{n}}));{{c}_{n}})] \\ 
\end{split}
\end{equation}

$\mathcal{K}_{f\to c}(\cdot)$ and $\mathcal{K}_{c\to t}(\cdot)$ are optimized jointly by introducing a hyperparameter $\lambda$ to balance their respective loss functions. While $0< \lambda < 1$ is a hyperparameter, $\mathscr{L}_{aux}$ serves as an auxiliary task to guide the learning process. We analyze the loss function under joint training: when $\lambda = 0$, the model reduces to a standard architecture without a concept bottleneck.

The three proposed concept bottleneck-based methods each emphasize different aspects of interpretability and performance. Therefore, in subsequent experiments, we conducted comparative analyses across all three schemes to identify the most effective architecture for PolSAR image classification. Additionally, we performed a hyperparameter search to determine the optimal value of the regularization coefficient $\lambda$. An overview of the entire proposed interpretable PolSAR image classification framework is illustrated in Fig. \ref{fig:model}.

\section{Experiment}
\subsection{Datasets Description and Experimental Setup}
To demonstrate that the proposed PaCBM framework achieves high interpretability while maintaining competitive classification accuracy, we conducted comparisons against four CNN-based methods (SF-CNN\cite{sfcnn}, PSENet\cite{pse}, CBAM\cite{cbam}, and MCFCNN\cite{MMcnn}), one Transformer-based method (ViT\cite{Dong}), two hybrid approaches (HyCVNet\cite{HybridCVNet} and PolFormer\cite{PolFormer}), and the interpretable model ProtoPNet\cite{protopnet}.

To further assess the effectiveness of each component within PaCBM, we performed an ablation study comparing the full PaCBM with the baseline CBM model and evaluated the impact of replacing the MLP with KAN to verify KAN’s advantage in mapping abstract concepts to class labels. We also conducted a hyperparameter search to determine the optimal value of the regularization coefficient $\lambda$. Finally, we demonstrateed the concept prediction capability of PaCBM through representative examples and validated its interpretability by reconstructing the functional expressions of selected KAN nodes. In addition, we illustrated the intervention capability of CBM by manually correcting concept predictions to achieve improved final classification outcomes.

All experiments were conducted on an NVIDIA GeForce GTX 2080 Ti GPU and the work was validated using three datasets: GF3 San Francisco (GF3-SF), RS2 San Francisco (RS2-SF)\cite{polSF} and  Oberpfaffenhofe\cite{Ob}. The details of the dataset are as follows:

\textit{1) GF3 San Francisco dataset:} The dataset is derived from the GF-3 satellite, which is mainly used for PolSAR image processing, classification and target detection. The dataset is an C-band full polarimetric image of San Francisco. The size of this image is 2304×2912, with 6 categories marked: mountain, water, vegetation, high-density urban areas, low-density urban areas and developed.

\textit{2) RS2 San Francisco dataset:} The dataset is derived from the Radarsat-2 satellite, which provides a C-band full polarimetric and high-resolution PolSAR image data of San Francisco. The dimensions of this image are 1800 × 1380 pixels with 5 categories marked: water, vegetation, high-density urban areas, low-density urban areas, and developed. 

\textit{3) Oberpfaffenhofe dataset:} The dataset is acquired by the ESAR airborne platform in 2002 in the L-band located on a German area. The dimensions of this image are 1300 × 1200 pixels with a spatial resolution size of 3.0 × 2.2m. It contains three categories: built-up area, wood land, open area.

In the sample generation process, to minimize interference from the surrounding neighborhood of the labeled anchor pixels, we set the sample patch size to 15×15 pixels. During the experiments, 2,000 samples were randomly selected from each category to form the training set, while an additional 500 samples from the remaining data were used for validation. In order to maintain generality, we did not use complex data augmentation methods, including commonly used PTD-based augmentation methods in the PolSAR image classification, but instead used the most basic PolSAR data as input, which consisted of the 9-dimensional PolSAR initial form \cite{high_dim}: \( T_{11} \), \( T_{22} \), \( T_{33} \), \( \text{Re}[T_{12}] \), \( \text{Im}[T_{12}] \), \( \text{Re}[T_{13}] \), \( \text{Im}[T_{13}] \), \( \text{Re}[T_{23}] \), and \( \text{Im}[T_{23}] \). In our experiment, all models were trained for 100 epochs with the batch size of 256. The Adam optimizer was used with the initial learning rate set to 0.001. 

Through the search of hyperparameters, the KAN network used a grid size of 7 and splines order of 3 to balance flexibility and efficiency. A two-layer KAN was employed for mapping concepts to categories, with 16 nodes in the hidden layer. The spline scale was set to 1.0 with independent scaling enabled for adaptive adjustment. Meanwhile, the lambda hyperparameter, used to balance the losses of the two parts, was set to 0.7. These settings ensured stable and effective training.
\begin{table*}[t]
    \centering
    \renewcommand{\arraystretch}{1.5}
    \caption{The classification performance of different methods on GF3-SF dataset. (\%) PaCBM-i, PaCBM-s, and PaCBM-j correspond to the independent, sequential, and joint training strategies, respectively.}
    \begin{tabular*}{\hsize}{@{\extracolsep{\fill}}ccccccccccccccc}
        \toprule
        \textbf{Class} & SF-CNN & PSENet & ViT & CBAM & ProtoPNet & MCFCNN & PolFormer & HyCVNet & PaCBM-i & PaCBM-s & PaCBM-j \\
        \midrule
        Mountain   & 73.29 & 79.28 & 74.06 & 83.08 & 81.42 & 86.95 & 86.75 & 79.17 & 86.94 & 86.58 & 90.16 \\
        Water      & 99.69 & 99.80 & 99.77 & 99.83 & 99.89 & 99.79 & 99.80 & 99.73 & 99.76 & 99.76 & 99.72 \\
        Vegetation & 62.38 & 66.22 & 69.83 & 75.00 & 83.01 & 80.80 & 77.04 & 89.07 & 81.14 & 81.92 & 76.71 \\
        HD-Urban   & 88.65 & 89.51 & 89.17 & 88.94 & 88.78 & 91.34 & 91.43 & 89.75 & 92.50 & 92.40 & 94.42 \\
        LD-Urban   & 93.29 & 95.27 & 96.19 & 95.33 & 95.31 & 95.57 & 96.54 & 96.54 & 97.08 & 96.93 & 95.32 \\
        Developed  & 90.88 & 93.85 & 92.42 & 90.95 & 94.64 & 95.21 & 95.14 & 92.56 & 95.48 & 95.34 & 96.33 \\
        \midrule
        AA         & 84.70 & 87.32 & 86.91 & 88.85 & 90.51 & 91.61 & 91.12 & 91.14 & 92.15 & \textbf{92.16} & 92.11 \\
        OA         & 92.38 & 93.47 & 93.59 & 94.25 & 95.06 & 95.49 & 95.26 & 95.69 & 95.88 & \textbf{95.90} & 95.73 \\
        Kappa      & 88.07 & 89.78 & 89.97 & 91.00 & 92.26 & 92.93 & 92.57 & 93.25 & 93.54 & \textbf{93.57} & 93.31 \\
        \bottomrule
    \end{tabular*}
    \label{GF3-comparative}
\end{table*}
\begin{figure*}[t]
    \centering
    \subfloat[]{
        \includegraphics[width=0.22\textwidth]{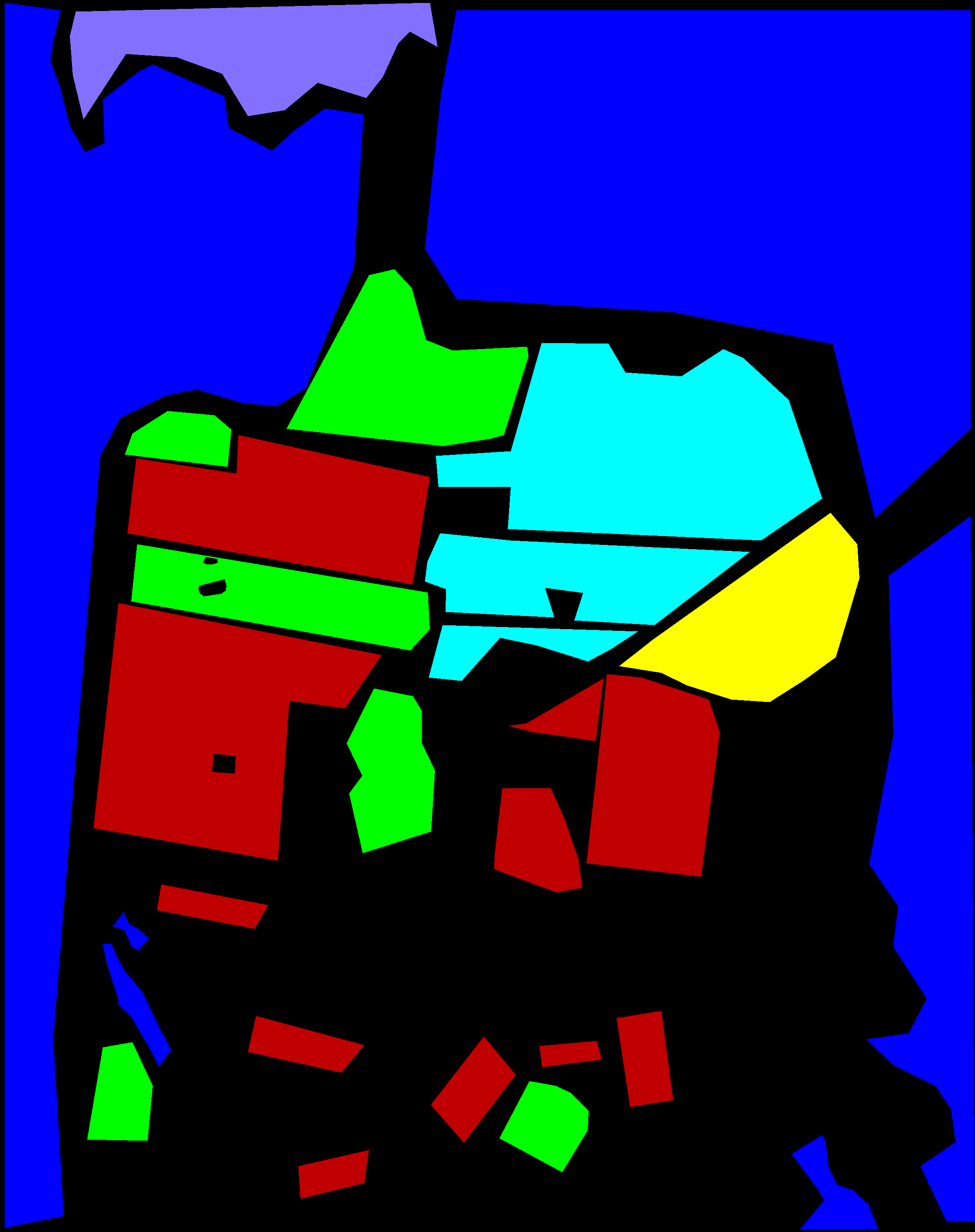}}~~\hfill~
    \subfloat[]{
        \includegraphics[width=0.22\textwidth]{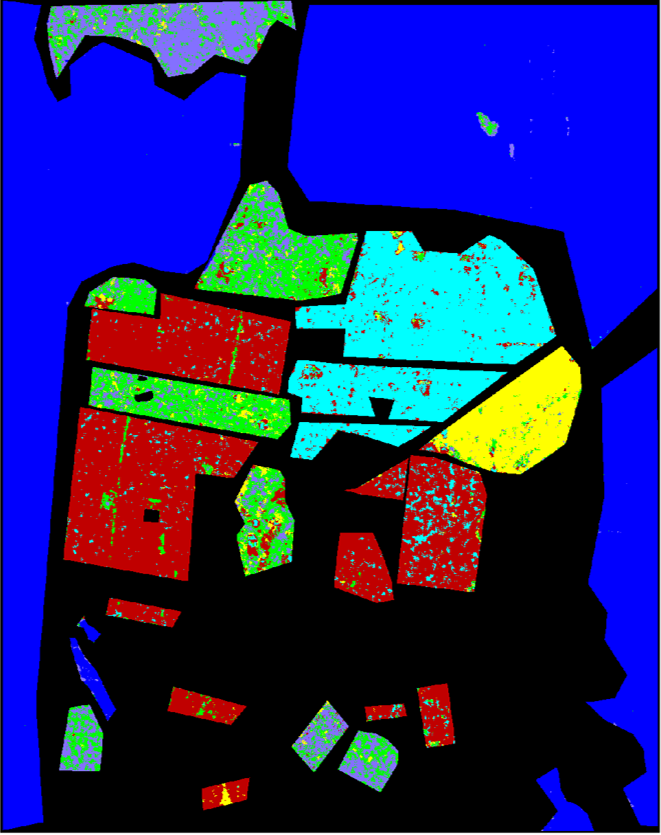}}~~\hfill~
    \subfloat[]{
        \includegraphics[width=0.22\textwidth]{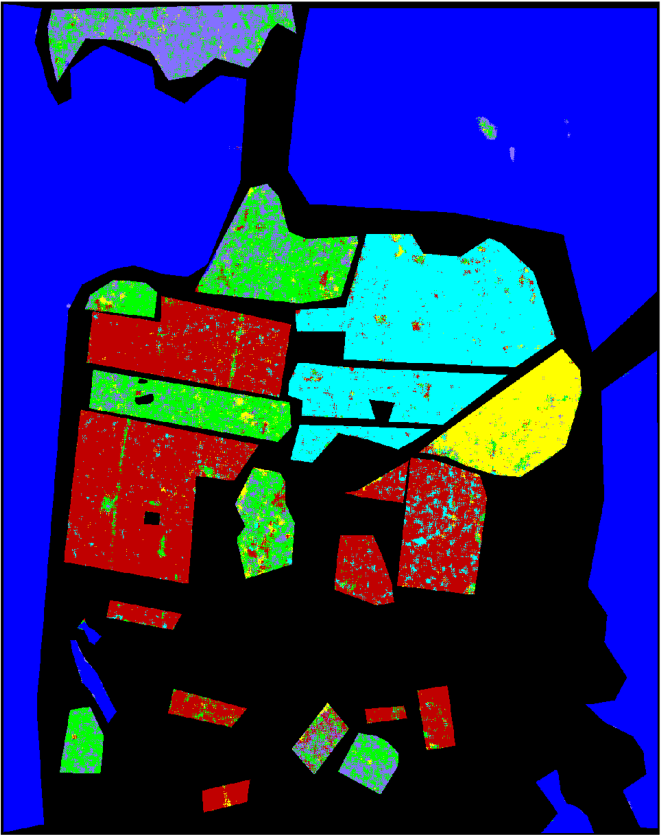}}~~\hfill~
    \subfloat[]{
        \includegraphics[width=0.22\textwidth]{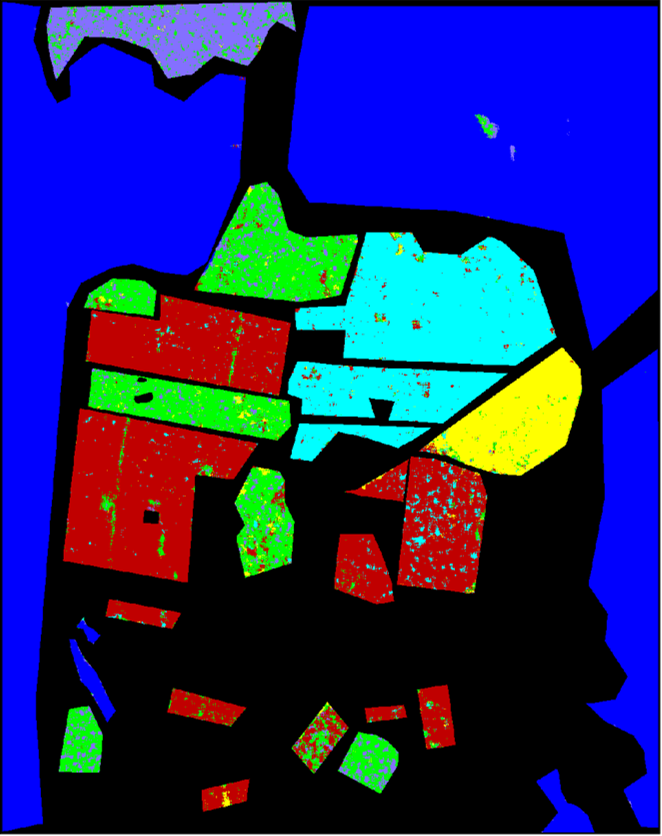}}
    \caption{Classification results of the whole map on GF3 San Francisco dataset. (a) GF3 label. (b) Result of SF-CNN. (c) Result of ViT. (d) Result of PaCBM.}
    \label{mapgf}
\end{figure*}
\begin{table*}[t]
    \centering
    \renewcommand{\arraystretch}{1.6}
    \caption{The classification performance of different methods on RS2-SF dataset. (\%) PaCBM-i, PaCBM-s, and PaCBM-j correspond to the independent, sequential, and joint training strategies, respectively.}
    \begin{tabular*}{\hsize}{@{\extracolsep{\fill}}ccccccccccccc}
        \toprule
        \textbf{Class} & SF-CNN & PSENet & ViT & CBAM & ProtoPNet & MCFCNN & PolFormer & HyCVNet & PaCBM-i & PaCBM-s & PaCBM-j \\
        \midrule
        Water & 99.94 & 99.95 & 99.96 & 99.97 & 99.94 & 99.99 & 99.99 & 99.99 & 99.97 & 99.97 & 99.90 \\
        Vegetation & 87.33 & 90.92 & 90.20 & 91.76 & 94.43 & 96.12 & 93.97 & 96.84 & 96.21 & 96.04 & 93.35 \\
        HD-Urban & 90.10 & 91.03 & 86.90 & 90.27 & 91.63 & 91.76 & 93.80 & 96.13 & 94.57 & 94.94 & 97.77 \\
        LD-Urban & 87.71 & 89.47 & 94.79 & 92.34 & 90.99 & 93.04 & 93.66 & 97.41 & 97.74 & 97.74 & 96.98 \\
        Developed & 91.80 & 93.38 & 94.32 & 96.72 & 96.36 & 98.23 & 97.34 & 93.56 & 97.29 & 97.38 & 98.39 \\
        \midrule
        AA & 91.38 & 92.95 & 93.23 & 94.21 & 94.67 & 95.83 & 95.75 & 96.79 & 97.16 & 97.21 & \textbf{97.28} \\
        OA & 93.95 & 94.98 & 94.97 & 95.57 & 95.95 & 96.24 & 96.81 & 98.04 & 97.92 & 97.97 & \textbf{98.06}\\
        Kappa & 91.39 & 92.86 & 92.85 & 93.70 & 94.23 & 94.65 & 95.47 & 97.20 & 97.04 & 97.11 & \textbf{97.23} \\
        \bottomrule
    \end{tabular*}
    \label{RS2-comparative}
\end{table*}
\begin{figure*}[t]
    \centering
    \subfloat[]{
        \includegraphics[width=0.22\textwidth]{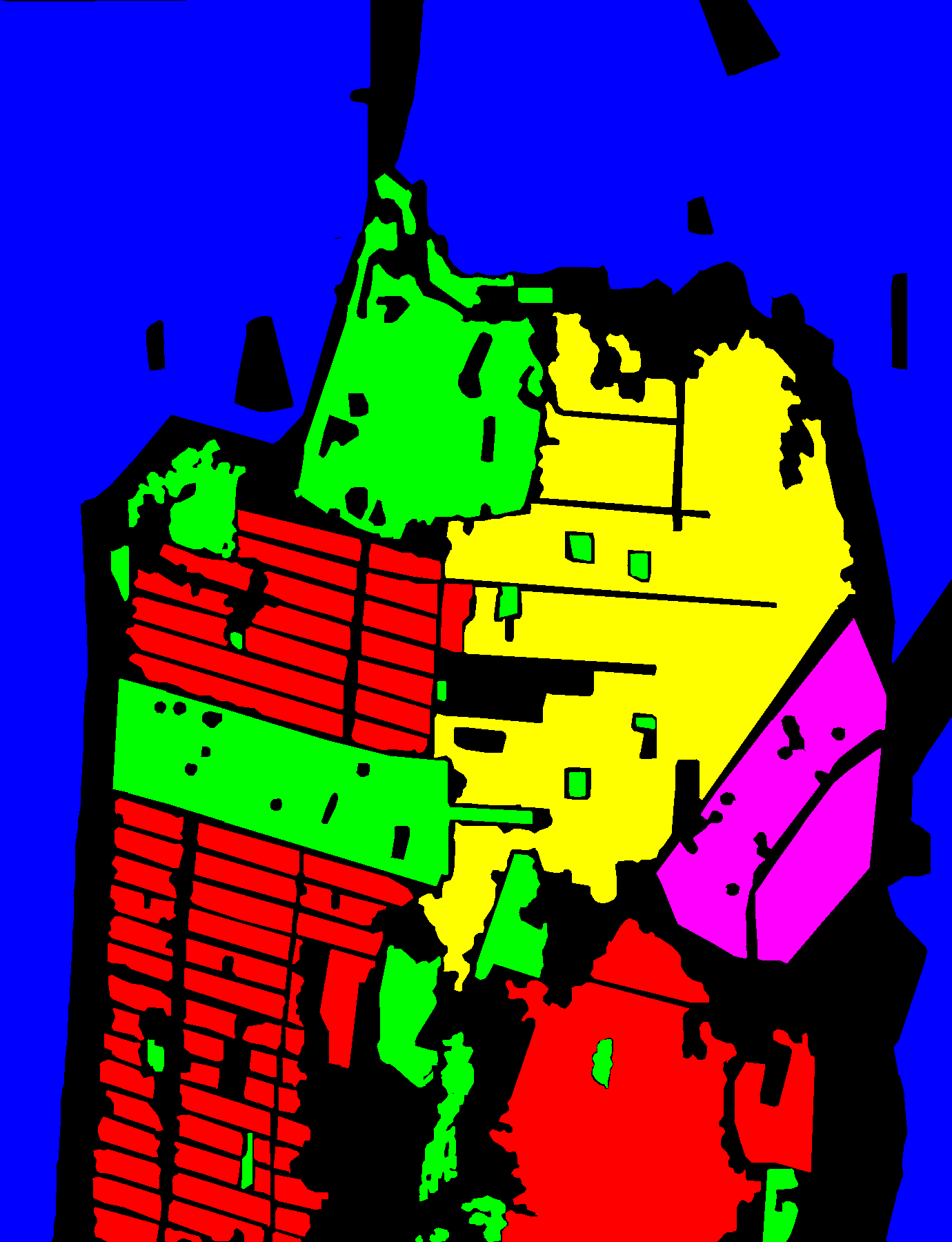}}~~\hfill~
    \subfloat[]{
        \includegraphics[width=0.22\textwidth]{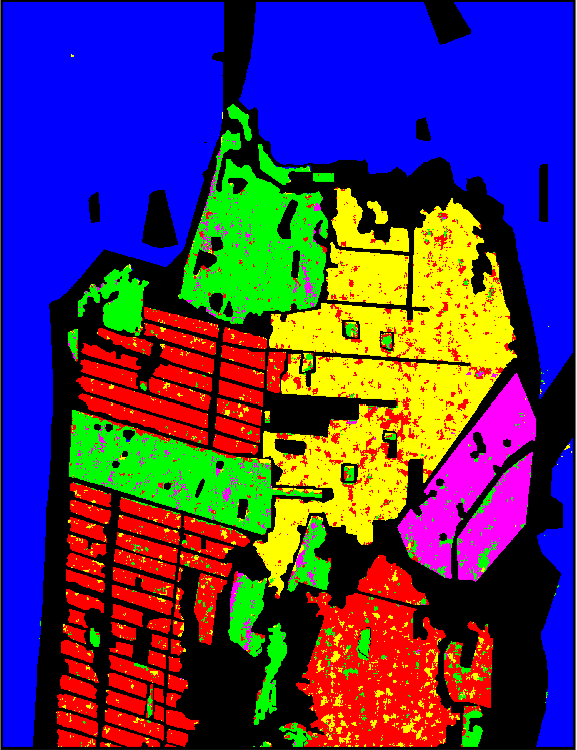}}~~\hfill~
    \subfloat[]{
        \includegraphics[width=0.22\textwidth]{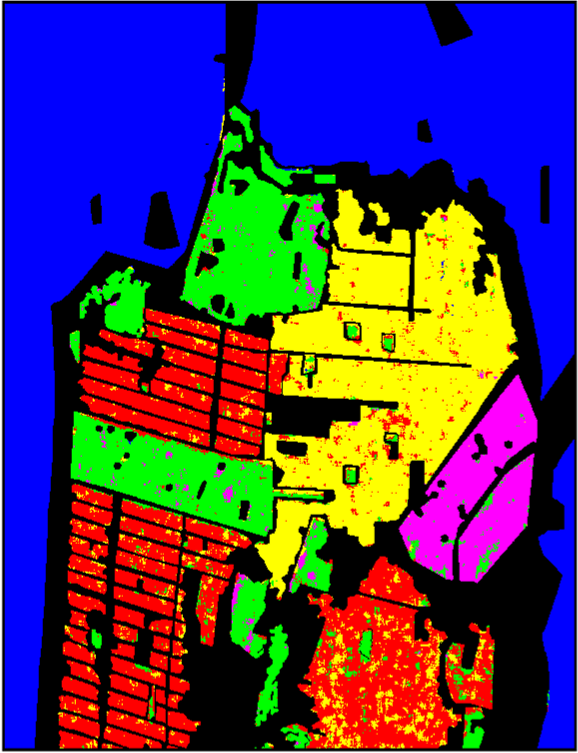}}~~\hfill~
    \subfloat[]{
        \includegraphics[width=0.22\textwidth]{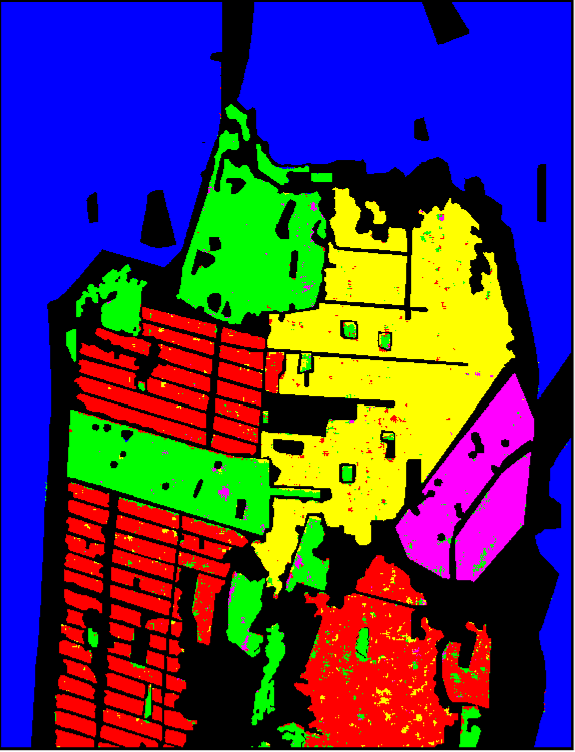}}

    \caption{Classification results of the whole map on RS2 San Francisco dataset. (a) RS2 label. (b) Result of SF-CNN. (c) Result of ViT. (d) Result of PaCBM.}
    \label{maprs}
\end{figure*}
There are three kinds of evaluation criteria including average accuracy (AA), Kappa coefficients (Kappa), and overall accuracy (OA) in the experiments. Among them, OA and AA  can measure the general performance of algorithms and Kappa is more important in the case of imbalanced data classification.

\subsection{Experiment Results and Analysis}
This section comprises four experimental components. First, contrast experiments were conducted on three benchmark datasets to evaluate classification performance across whole map. The results demonstrated that the proposed method achieves competitive accuracy while preserving model interpretability. Second, to verify the effectiveness of the proposed PaCBM-KAN framework, ablation studies were performed using the RS2-SF dataset. Finally, we explored the optimal settings for the hyperparameter $lambda$ and the structural configuration of the KAN network.

\subsubsection{Contrast Experiments}
Although the proposed PaCBM was primarily designed to enhance interpretability in the classification process, it nevertheless achieved competitive performance compared to state-of-the-art methods. The superior performance of PaCBM in terms of OA, AA, and Kappa could be attributed to two main factors. First, PaCBM integrated a powerful dual-branch encoder that combined CNN and ViT, enabling it to effectively capture both local spatial features and global contextual information. Second, the model employed a parallel structure that mitigated the information loss typically encountered in traditional CBM. Moreover, the inclusion of concept labels introduced additional supervision information, which not only enhanced the interpretability of the model but also facilitated more robust and discriminative feature extraction. In the following, we provide a detailed analysis of the results obtained on the three datasets.

For the experiments on the GF3-SF dataset, the results of AA, OA, and Kappa are reported in Table \ref{GF3-comparative}, and the whole classification maps are shown in Fig. \ref{mapgf}. All three training variants of the proposed model outperform the eight comparison methods. Among them, PaCBM-s has achieved the best performance, with AA, OA, and Kappa reaching 92.16\%, 95.90\%, and 93.57\%, respectively. Compared to HyCBNet, the best-performing method among the baselines, PaCBM-s achieves an improvement of 1.02\% in AA. Compared to another interpretable method, ProtoPNet, our proposed approach achieves improvements of 1.65\%, 0.84\%, and 1.31\% in AA, OA, and Kappa, respectively. In terms of per-class classification accuracy, our method demonstrate clear advantages in distinguishing mountain, high-density urban, and low-density urban areas. This is further supported by the whole map classification results, where the colors representing these classes appeared more distinct and pure.

For the experiments on the RS2-SF dataset, the results of AA, OA, and Kappa are reported in Table \ref{RS2-comparative}, and the whole classification maps are shown in Fig. \ref{maprs}. The experimental results on this dataset are consistent with the observations on GF3-SF: all three training variants of the proposed PaCBM outperform the comparison methods in terms of AA. However, in terms of OA and Kappa, the independent and sequential training methods are slightly inferior to HyCVNet. In this case, PaCBM-j has achieved the best performance, with AA, OA, and Kappa reaching 97.28\%, 98.06\%, and 97.23\%, respectively. Compared to another interpretable method, ProtoPNet, our proposed approach achieves improvements of 2.61\%, 2.11\%, and 3.00\% in AA, OA, and Kappa, respectively. From the whole map classification results, PaCBM demonstrate a significant advantage in distinguishing Developed and High-Density Urban areas, where the colors representing these classes appeared more distinct and pure.

\begin{figure*}[t]
    \centering
    \subfloat[]{\label{fig:ob-label}\includegraphics[width=0.22\textwidth]{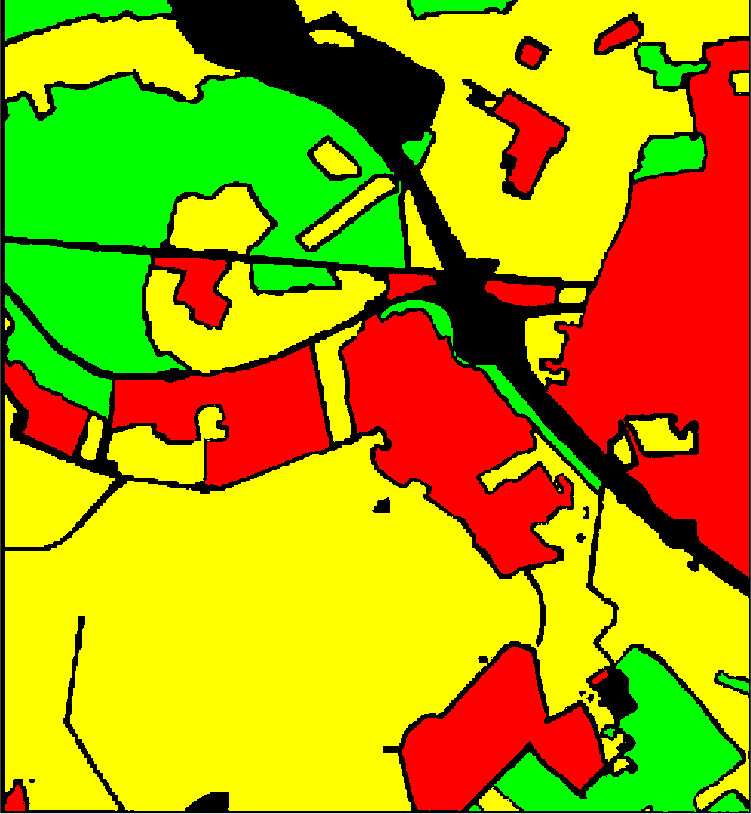}}%
    \hfill
    \subfloat[]{\label{fig:ob-sf}\includegraphics[width=0.22\textwidth]{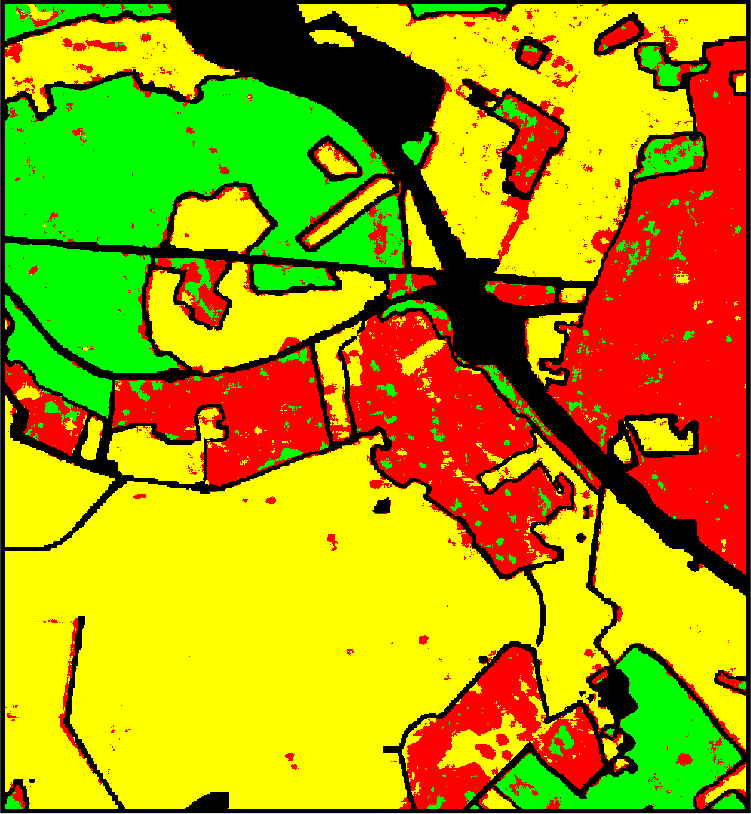}}%
    \hfill
    \subfloat[]{\label{fig:ob-vit}\includegraphics[width=0.22\textwidth]{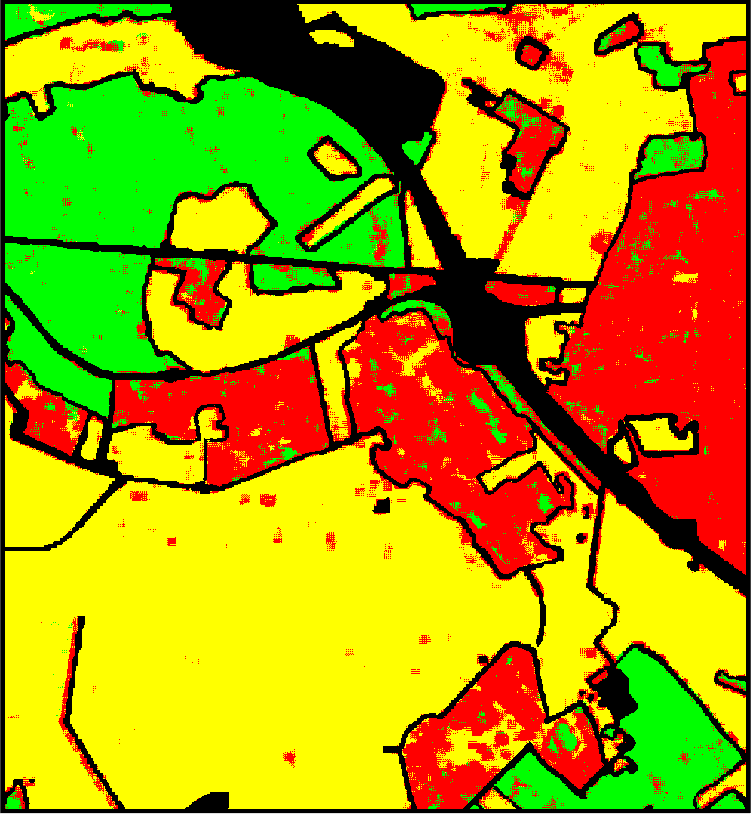}}%
    \hfill
    \subfloat[]{\label{fig:ob-pacbm}\includegraphics[width=0.22\textwidth]{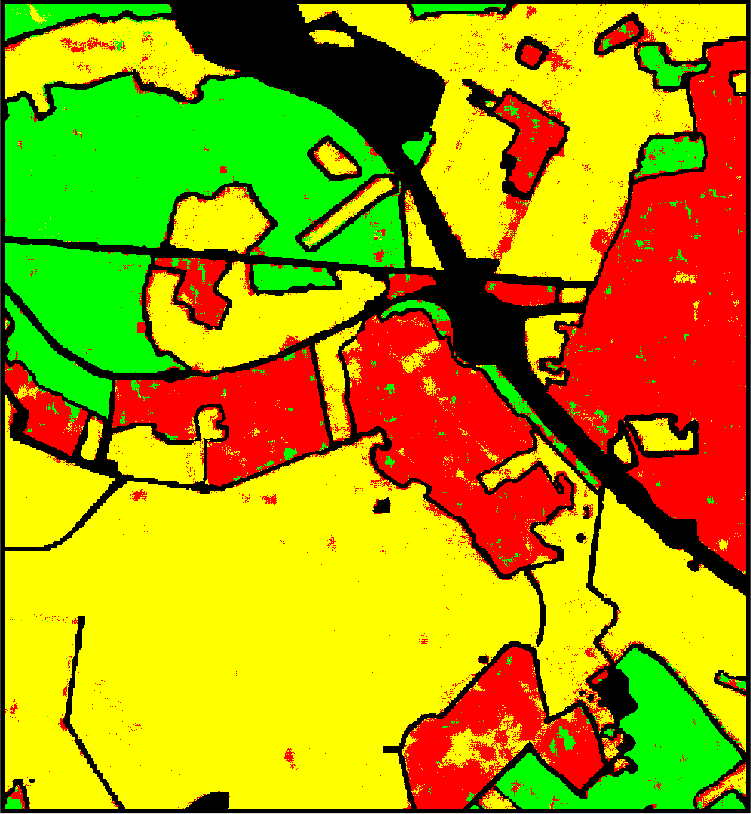}}%
    
    \caption{Classification results of the whole map on Oberpfaffenhofen dataset.  (a) Oberpfaffenhofen label. (b) Result of SF-CNN. (c) Result of ViT. (d) Result of PaCBM.}
    \label{obmap}
\end{figure*}

For the experiments on the Oberpfaffenhofen dataset, the results of AA, OA, and Kappa are reported in Table \ref{Ob-comparative}, while the whole classification maps are shown in Fig. \ref{obmap}. The experimental results demonstrate that the proposed PaCBM variants consistently outperformed most comparison methods across all evaluation metrics. Among these variants, PaCBM-s has achieved the best overall performance, with AA, OA, and Kappa values reaching 95.51\%, 95.58\%, and 92.49\%, respectively. Compared to HyCBNet, which represent the best-performing baseline method, PaCBM-s achieves a improvement of 0.22\% in AA. Among the different land cover classes, all PaCBM variants perform particularly strong performance on Wood Land classification, with PaCBM-j achieving 98.07\% accuracy. From the whole map classification results, PaCBM demonstrate a significant advantage in distinguishing Wood Land and Open Area, where the colors representing these classes appeared more distinct and pure.

In summary, experiments on both datasets demonstrated that, although our method was primarily motivated by enhancing the interpretability of DL–based PolSAR image classification models, the proposed PaCBM also achieved competitive classification performance on par with state-of-the-art methods.

\subsubsection{Ablation Experiments}
To thoroughly evaluate the advantages of our proposed PaCBM over the original CBM, as well as the superiority of KAN compared to MLP in nonlinear modeling, we conducted ablation experiments on the RS2-SF dataset. The baseline model consisted solely of the backbone network illustrated in Fig. \ref{fig:model}(b). The CBM followed the $x \to c \to y$ structure without incorporating the parallel branch. PaCBM with MLP employed an MLP to map concepts to labels, whereas PaCBM-j—the proposed approach—replaced the MLP with KAN. Both CBM and PaCBM variants were trained using the joint training strategy in these ablation studies. The results are summarized in Table \ref{rs2_ab}.
\begin{table}[h]
    \centering
    \renewcommand{\arraystretch}{1.5}
    \caption{Ablation study on RS2-SF dataset}
    \begin{tabular*}{\hsize}{@{\extracolsep{\fill}}ccccccccccccccc}
    \hline
    Method & Baseline & CBM & PaCBM with MLP & PaCBM with KAN \\
    \hline
    AA & 97.11 & 96.41 & 97.07 & 97.28 \\
    OA  & 97.86 & 97.75 & 97.85& 98.06 \\
    Kappa & 96.95 & 96.79 & 96.88 & 97.23 \\
    \hline
    \end{tabular*}
    \label{rs2_ab}
\end{table}

Compared to the baseline, CBM exhibits a decline across all three metrics, with AA experiencing a notable drop of 0.7\%. This indicates that the direct introduction of the CBL results in a certain degree of information loss, which inevitably impacts classification performance. This decline can be attributed to the inherent “information leakage” issues of CBM, as also identified in prior studies \cite{cbm_post-hoc,cbm_Editable}. In contrast, PaCBM with MLP—which introduced a parallel branch—mitigates these risks by preserving the independence between image feature information and concept representations. Consequently, it achieves performance close to the baseline, improving AA by 0.66\% compared to CBM. Finally, PaCBM-j, which replaced the MLP with KAN, further improves all three metrics by 0.21\%, 0.21\%, and 0.35\%, respectively, surpassing the baseline performance. These results demonstrate that KAN more effectively captures the nonlinear mapping between concepts and labels. In summary, the ablation study shows that the proposed PaCBM effectively alleviates information leakage, and that KAN not only enhances interpretability but is also better suited for modeling nonlinear relationships.
\begin{table*}[t]
    \centering
    \renewcommand{\arraystretch}{1.6}
    \caption{The classification performance of different methods on Oberpfaffenhofen dataset. (\%) PaCBM-i, PaCBM-s, and PaCBM-j correspond to the independent, sequential, and joint training strategies, respectively.}
    \begin{tabular*}{\hsize}{@{\extracolsep{\fill}}ccccccccccccc}
        \toprule
        \textbf{Class} & SF-CNN & PSENet & ViT & CBAM & ProtoPNet & MCFCNN & PolFormer & HyCVNet & PaCBM-i & PaCBM-s & PaCBM-j \\
        \midrule
        Built-up Area & 88.91 & 90.88 & 89.62 & 92.53 & 87.94 & 91.43 & 92.48 & 92.40 & 93.30 & 93.28 & 92.79 \\
        Wood Land & 93.45 & 95.79 & 94.75 & 95.62 & 96.77 & 93.88 & 96.61 & 97.54 & 97.15 & 97.14 & 98.07 \\
        Open Area & 95.72 & 95.82 & 94.64 & 95.94 & 97.84 & 96.78 & 96.47 & 95.93 & 96.04 & 96.08 & 95.28 \\
        \midrule
        AA & 92.69 & 94.17 & 93.01 & 94.70 & 94.18 & 94.03 & 95.19 & 95.29 & 95.50 &  \textbf{95.51} & 95.38 \\
        OA & 93.60 & 94.58 & 93.41 & 95.03 & 95.17 & 94.90 & 95.50 & 95.35 & 95.57 &  \textbf{95.58} & 95.50 \\
        Kappa & 89.11 & 90.80 & 88.84 & 91.56 & 91.73 & 91.31 & 92.34 & 92.12 & 92.47 &  \textbf{92.49} & 92.38 \\
        \bottomrule
    \end{tabular*}
    \label{Ob-comparative}
\end{table*}

\subsubsection{Discussions on Hyperparameter $\lambda$}
The hyperparameter $\lambda$ plays a crucial role in unlocking the performance of PaCBM. When $\lambda$ is set too high, the model tends to overemphasize the concept prediction task, which may come at the expense of final classification accuracy. Conversely, when $\lambda$ is too low, the guiding effect of the parallel branch on feature learning toward semantically meaningful concepts is weakened, limiting its benefits. 
\begin{figure}[h]
    \centering
    \includegraphics[width=1.0\linewidth]{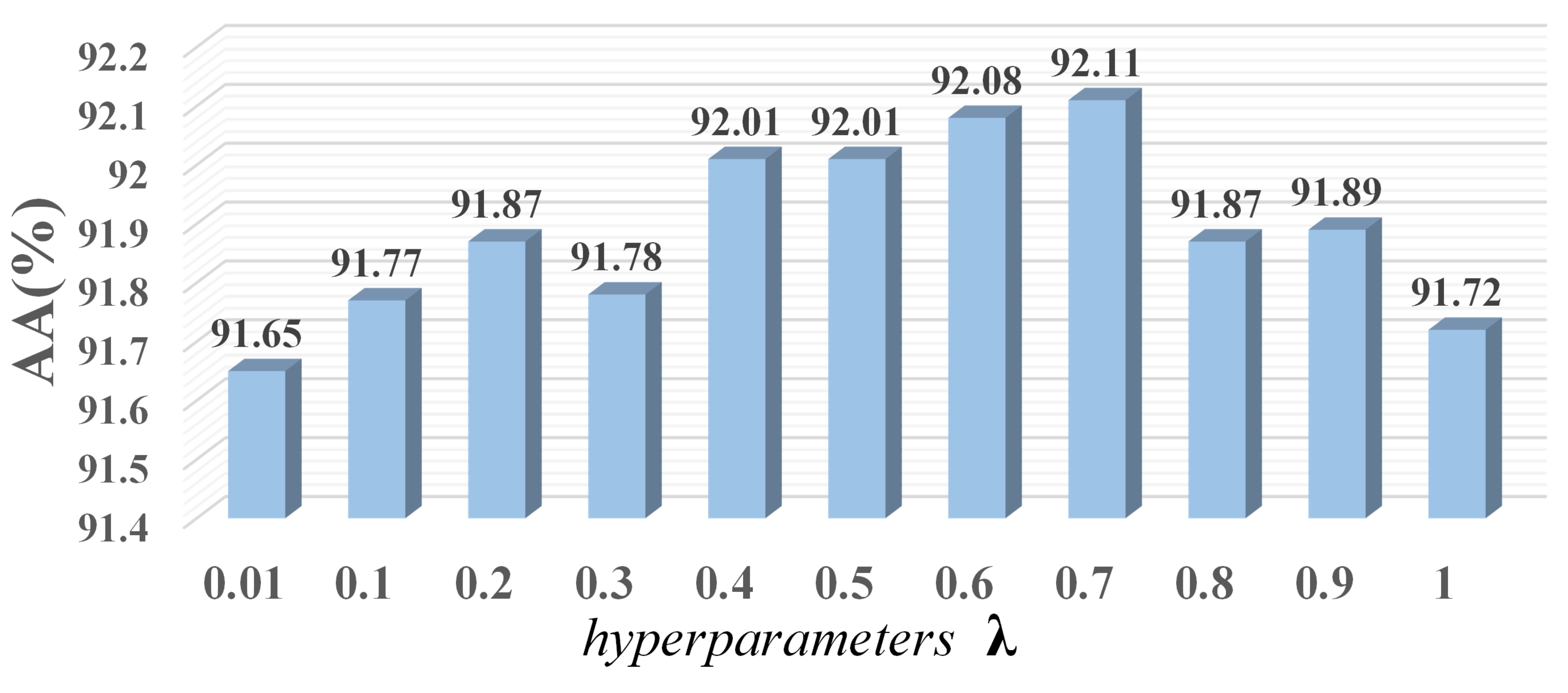}
    \caption{Effect of hyperparameter $\lambda$ on AA in the GF3-SF dataset}
    \label{fig:lamda}
\end{figure}

To identify the optimal balance, we conducted a hyperparameter search on the GF3-SF dataset, exploring $\lambda \in [0.01, 0.1, 0.2, 0.3,\ 0.4, 0.5, 0.6, 0.7, 0.8, 0.9, 1]$, using AA as the evaluation metric for model selection. The results of this search are presented in Fig. \ref{fig:lamda}. Based on the experimental results, the hyperparameter \(\lambda\) significantly influences the performance of PaCBM-joint. The observed trend shows that the AA initially increases with \(\lambda\) reaching a peak before declining as \(\lambda\) continues to grow. Relatively strong performance is maintained within the range \([0.4, 0.7]\), with \(\lambda = 0.7\) achieving the highest accuracy among all tested values.

\subsubsection{Discussions on KAN Network Structure}
\begin{figure}[h]
    \centering
    \includegraphics[width=1.0\linewidth]{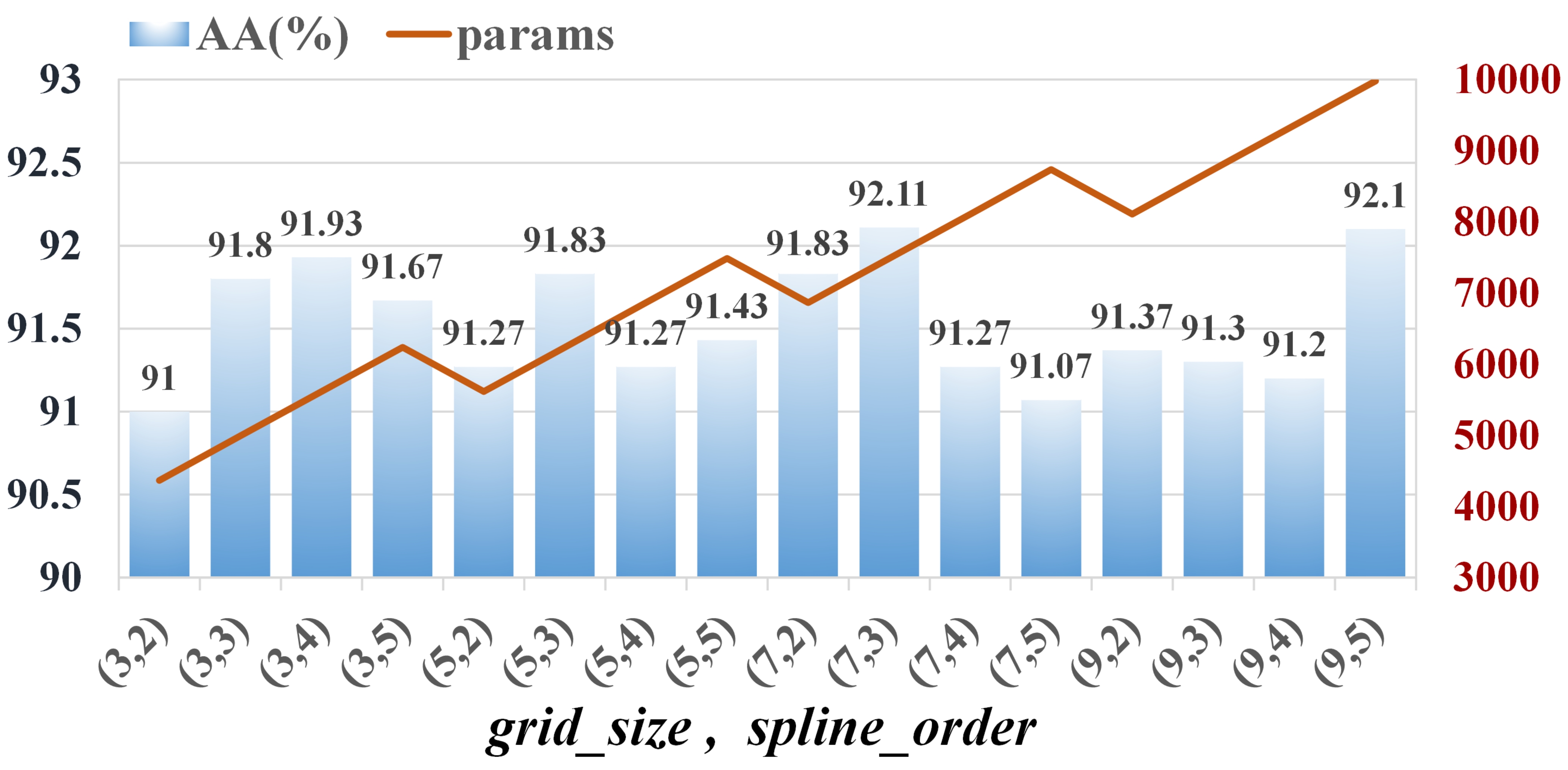}
    \caption{Impact of grid size and spline order on AA and params in KAN}
    \label{fig:kan_sousuo}
\end{figure}

To evaluate the influence of the hyperparameters grid size and spline order in the KAN architecture, we explored the combination of $grid\ size \in [3,5,7,9]$ and $spline\ order \in [2,3,4,5]$, taking into account both AA and params. As shown in Fig. \ref{fig:kan_sousuo}, increasing grid size and spline order generally enhances the AA; however, this improvement is accompanied by a substantial increase in the number of parameters. Among the tested configurations, the setting $(7, 3)$ achieves the highest AA of 92.11\% with a moderate parameter count of approximately 7,000, thereby offering a favorable balance between performance and model complexity. In comparison, although configurations such as $(9, 5)$ also yield comparable accuracy (92.10\%), they require a significantly higher number of parameters (around 10,000), rendering them less efficient. Based on this trade-off, the $(7, 3)$ configuration was selected as the optimal setting for our model, as it provides the best compromise between accuracy and computational efficiency.

\subsection{Interpretability Analysis}
This section demonstrates the strong interpretability of the proposed PaCBM model from three perspectives. First, using the concept labels constructed in our work, the model not only provides terrain surface category predictions for PolSAR samples but also offers interpretability by visualizing the predicted concept values, thereby unveiling the learned high-dimensional features. Second, thanks to KAN’s concise yet powerful nonlinear modeling capabilities, the model’s inference process can be traced and expressed as a sum of simple B-spline functions, enabling visualization of how concept labels influence the final prediction. Finally, by leveraging domain knowledge and existing interpretability tools, we can intervene on the predicted concept labels to adjust and refine the ultimate classification outcome.

\subsubsection{Concept Labels Prediction}
\begin{figure}[h]
    \centering
    \includegraphics[width=1.0\linewidth]{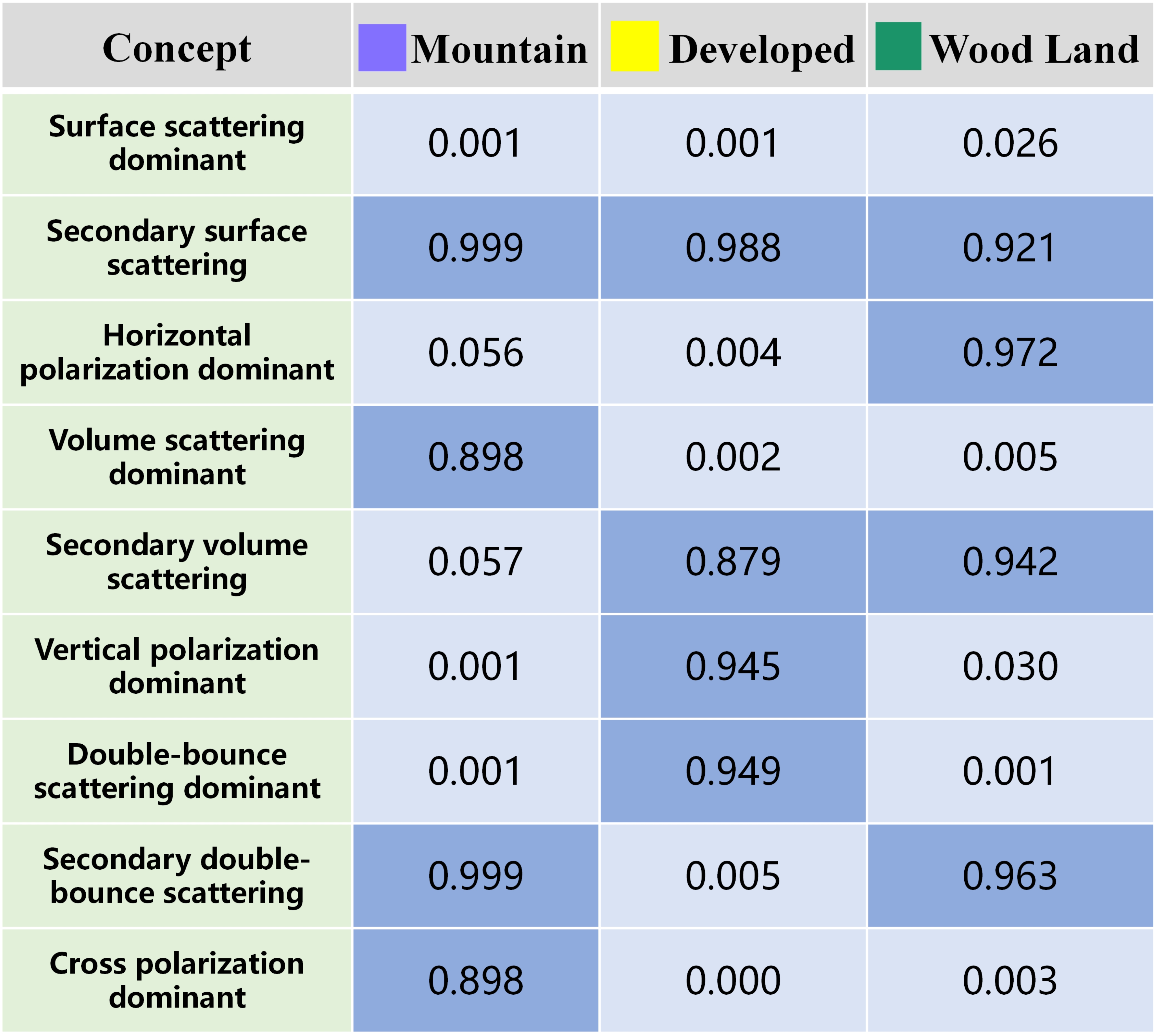}
    \caption{Some results of the concept prediction in PolSAR dataset. We selected one sample from each dataset: Mountain from GF3-SF, Developed from RS2-SF, and Wood Land from Oberpfaffenhofen, and visualized their predictions onto several concept labels. Dark colors represent predictions close to 1, and light colors represent predictions close to 0.}
    \label{fig:cbm_labels}
\end{figure}

To validate the strong correlation between concepts and target features, we selected one sample from each dataset: Mountain from GF3-SF, Developed from RS2-SF, and Wood Land from Oberpfaffenhofen. Their predictions on several concept labels were visualized, as shown in Fig. \ref{fig:cbm_labels}.

From both common sense and electromagnetic characteristics, a mountain sample is expected to be dominated by volume scattering due to its complex and variable structure. Our experimental results confirm this, with the volume scattering dominant concept scoring 0.898. Additionally, secondary surface scattering and secondary double-bounce scattering both receive high predicted values of 0.999, which may be attributed to the presence of forest structures on the mountain. The cross-polarization dominant concept is also predicted with a high value of 0.898, consistent with rough asymmetric surfaces typical of mountainous terrain. For the developed area sample, double-bounce scattering dominant is expected to be prominent because of the numerous dihedral angle structures found in urban buildings. Secondary volume scattering likely arises from complex trees and urban facilities, while vertical polarization dominant reflects the prevalence of vertical structures, consistent with the building architecture. Our model predicts values of 0.949, 0.879, and 0.945 for double-bounce scattering dominant, secondary volume scattering, and vertical polarization dominant, respectively. Similarly, for woodland, its complex structure—such as volume scattering from tree canopies and possible dihedral scattering from tree stumps—leads to high predictions for secondary volume scattering and secondary double-bounce scattering, with scores of 0.942 and  0.963.

These three examples demonstrate that PaCBM enables precise conceptual analysis of previously unseen samples. When a sample is input into the model, its polarization concepts are first extracted via the CBL, yielding high scores for concepts relevant to the target features and low scores for irrelevant ones. These polarization concepts serve as accurate, physically meaningful intermediate representations that bridge the gap between raw input features and the final classification. This concept-to-label prediction process closely mirrors human reasoning, thus providing a solid foundation for interpretable PolSAR image classification.

\begin{figure*}[ht]
    \centering
    \includegraphics[width=1.0\linewidth]{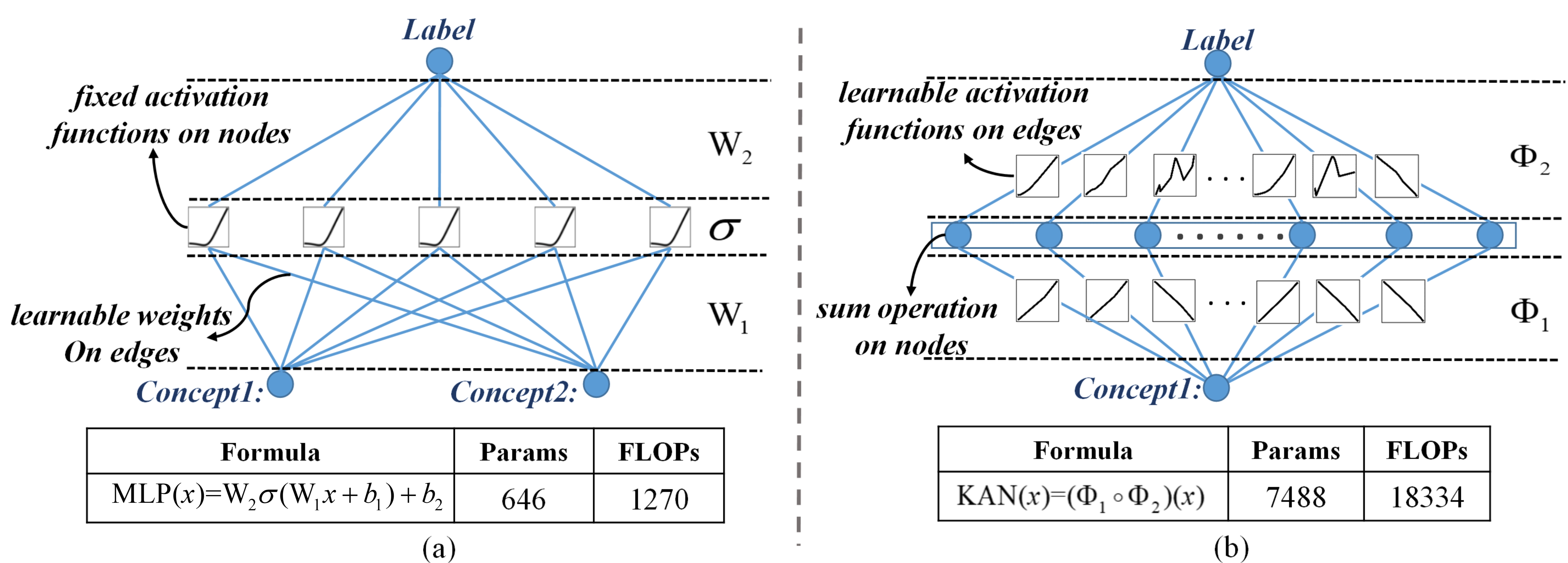}
    \caption{Visualization of Internal Computation and Comparison between KAN and MLP. (a) MLP performs mapping via fixed activation functions combined with learnable weights on the edges. (b) KAN achieves mapping by using learnable activation functions and summation on the edges.}
    \label{fig:kan}
\end{figure*}
\begin{table*}[t]
\centering
\caption{Symbolic Formula Representations for Concept-to-Label Mapping of Land Cover Categories in the GF3-SF Dataset}
\begin{tabular}{c>{\raggedright\arraybackslash}p{\dimexpr\textwidth-3cm\relax}}
\specialrule{1.2pt}{0pt}{0pt}
\rule{0pt}{13pt}\textbf{\small Category} & \textbf{\normalsize Symbolic Formula} \\
\midrule
 \textbf{\footnotesize Mountain} & $0.01c_1 - 0.31\sin(0.96c_2 + 8.39) - 0.05(-0.74c_3 - 1)^2 + 0.04c_4 - 0.04c_5 + 0.03\sin(1.48c_6 - 1.39) + 0.04c_8 + 0.05c_9 + 0.13c_{10} + 0.16\sin(0.89c_{11} + 5.19) + 0.22c_{12} - 0.08c_{13} - 0.12c_{14} + 0.26c_{15} + 0.03c_{16} - 0.06c_{17} - 0.10c_{18} - 0.01c_{19} - 0.03(-c_{20} - 0.72)^2 + 0.14c_{23} - 0.07c_{24} + 0.22e^{0.52c_{26}} - 0.08c_{27} + 0.01c_{28} + 0.05c_{29} + 0.01c_{30} + 0.06c_{31} - 0.06c_{32} - 0.04c_{33} + 0.28$
\\
\midrule
 \textbf{\footnotesize Water} & $
0.03c_2 - 0.05c_3 + 0.02c_4 + 0.01c_5 - 0.01c_6 + 0.07c_8 + 0.16c_9 - 0.03c_{10} + 0.05(-0.67c_{11} - 1)^2 + 0.01c_{12} + 0.12c_{13} - 0.04c_{16} + 0.01c_{17} + 0.23c_{18} - 0.04c_{19} + 0.05c_{20} - 0.01c_{23} + 0.16c_{24} + 0.06c_{25} - 0.04c_{26} + 0.02(-0.89c_{27} - 1)^2 - 0.04c_{28} - 0.02(-0.85c_{29} - 1)^2 + 0.05c_{30} + 0.10c_{31} - 0.04(-0.8c_{32} - 1)^2 - 0.06c_{33} 
$
\\
\midrule
 \textbf{\footnotesize Vegetation} & $
0.22(-0.81c_1 - 1)^2 + 0.08c_2 + 0.61c_3 - 0.34 e^{0.56c_4} + 0.72 \sin(1.7 c_5 - 1.39) - 0.21 c_6 + 0.07 (-0.82 c_8 - 1)^2  - 0.15 c_9 + 0.67 c_{10} + 0.48 c_{11} + 0.02 (-c_{12} - 0.96)^2 - 0.56 c_{13} - 0.82 c_{14} + 0.22 e^{0.57 c_{15}} - 0.22 c_{16} + 0.83 \cos(1.34 c_{17} - 5.79)- 0.16 c_{18} - 0.03 c_{23} - 0.09 c_{24} - 0.05 c_{25} + 0.15 c_{26} - 0.65 \log(6.63 - 3.21 c_{27}) + 0.32 c_{28} + 0.97 \sin(0.64 c_{29} + 5.22) - 0.35 \sin(2.94 c_{30} + 1.64) - 0.07 c_{31} + 0.32 c_{32} + 1.72$
 \\
\midrule
 \textbf{\footnotesize High-Density Urban} &  $
0.04 c_1 + 0.12 c_2 - 0.28 c_3 - 0.28 c_4 + 0.33 c_5 - 0.04 (0.4 - c_6)^2 + 0.24 \cos(2.12 c_8 - 2.96) - 0.16 c_9 - 0.14 c_{10} - 0.19 \cos(1.36 c_{11} + 9.8) + 0.05 c_{12} - 0.06 c_{13} - 0.36 \sin(0.72 c_{14} - 4.21) + 0.16 c_{15} + 0.12 c_{16} - 1.04 \cos(1.13 c_{17} - 5.79) - 0.05 c_{19} - 0.23 c_{23} - 0.03 c_{24} + 0.05 c_{25} - 0.11 c_{26} - 0.07 c_{27} - 0.7 \sin(0.88 c_{28} + 5.21) - 0.06 c_{29} + 0.07 c_{31} - 0.05 c_{32} + 0.05 c_{33} + 0.73
$
\\
\midrule
 \textbf{\footnotesize Low-Density Urban} &  $
0.1 c_1 + 0.11 e^{0.54 c_2} - 0.16 \sin(1.13 c_3 + 5.21) - 0.05 (-0.74 c_4 - 1)^2 - 0.08 c_5 - 0.03 (-c_6 - 0.35)^2 - 0.03 c_8 - 0.06 (0.25 - c_9)^2 - 0.06 c_{10} - 0.19 c_{11} + 0.01 c_{12} - 0.32 \sin(1.25 c_{13} + 8.19) + 0.14 c_{14} - 0.12 \sin(1.23 c_{15} - 7.4) + 0.21 e^{0.57 c_{16}} - 0.16 c_{17} - 0.09 (0.45 - c_{18})^2 + 0.42 c_{19} + 0.28 (0.89 - c_{20})^2 + 0.11 c_{23} + 0.01 c_{24} - 0.02 c_{25} + 0.01 c_{26} - 0.86 \cos(0.68 c_{28} - 5.83) - 0.04 c_{29} - 0.11 c_{30} - 0.06 c_{31} + 0.01 c_{32} + 0.03 c_{33} + 0.58
$
\\
\midrule
 \textbf{\footnotesize Developed} & $
-0.5 c_1 - 0.41 c_2 - 0.18 \sin(0.96 c_3 - 7.4) + 0.76 c_4 + 1.78 \sin(0.82 c_5 + 8.4) + 0.3 c_6 - 0.4 c_8 + 0.25 c_9 + 0.83 \sin(0.86 c_{10} - 4.18) - 0.38 c_{11} - 0.34 c_{12} + 0.39 c_{13} + 1.86 \sin(0.7 c_{14} + 5.24) - 0.38 c_{15} - 0.05 \sin(1.43 c_{16} - 7.61) - 0.03 c_{17} + 0.15 c_{18} + 0.23 c_{19} + 0.49 c_{20} + 0.44 c_{23} + 0.08 c_{24} + 0.27 c_{25} - 0.09 (-0.98 c_{27} - 1)^2 - 0.1 c_{28} - 0.57 \sin(0.98 c_{29} + 5.21) - 0.93 c_{30} - \frac{0.01}{0.18 - 0.55 c_{31}} - 0.01 \sin(2.28 c_{32} - 2.01) + 0.13 c_{33} - 1.17
$
 \\
\specialrule{1.2pt}{0pt}{0pt}
\end{tabular}
\label{formula}
\end{table*}

\subsubsection{Mapping from Concepts to Labels}
To gain deeper insights into the reasoning process of the model, we visualized the inference behavior of the KAN, as illustrated in Fig. \ref{fig:kan}. KAN demonstrates not only superior predictive performance in mapping abstract concepts to classification labels but, more importantly, markedly improved interpretability. This is primarily attributed to its use of learnable spline functions in place of fixed nonlinear activation functions (e.g., ReLU, Sigmoid). These spline-based components possess explicit mathematical formulations, thereby enabling a transparent and traceable inference pathway from input features to output predictions. This contrasts with the black-box nature of MLPs, wherein nested nonlinear transformations $\mathrm{MLP}(x) = \mathbf{W}_2 \, \sigma(\mathbf{W}_1 x + \mathbf{b}_1) + \mathbf{b}_2$ hinder interpretability. A well-established fact is that while KAN introduces interpretability, it also brings a significant increase in the number of parameters and computational complexity.

To illustrate the quantification process from concepts to labels, Table \ref{formula} presents symbolic formulas derived using KAN, representing the conceptual mapping from land cover semantics to classification labels. Each formula corresponds to a specific land cover category (e.g., Mountain, Water, Urban) and captures the underlying geometric and functional characteristics through interpretable mathematical expressions composed of basic operations (e.g., sine, power, polynomial terms). Concept prediction is performed via a formula-based approach, in which concept values serve as input variables and are individually processed using six predefined formulas. This process yields six corresponding logit scores, with the highest determining the predicted category. In contrast to the activation patterns and weight combinations in MLPs—which rarely produce compact or human-readable formulas—KAN enables each node and connection to encode a meaningful function. As a result, the network’s output can be directly expressed as a closed-form symbolic equation. This symbolic formulation not only enhances interpretability by revealing how input features relate to semantic categories but also facilitates downstream analytical tasks such as formula inspection, simplification, and domain-specific reasoning.

\subsubsection{Concept Intervention and Editing}
Concept intervention and editing serve as strong evidence for the interpretability of CBM, as they help to demystify the black-box behavior of neural networks. By enabling adjustments to intermediate concept representations, this mechanism improves model transparency and fosters greater user confidence in its decisions.
\begin{figure}[h]
    \centering
    \includegraphics[width=1.0\linewidth]{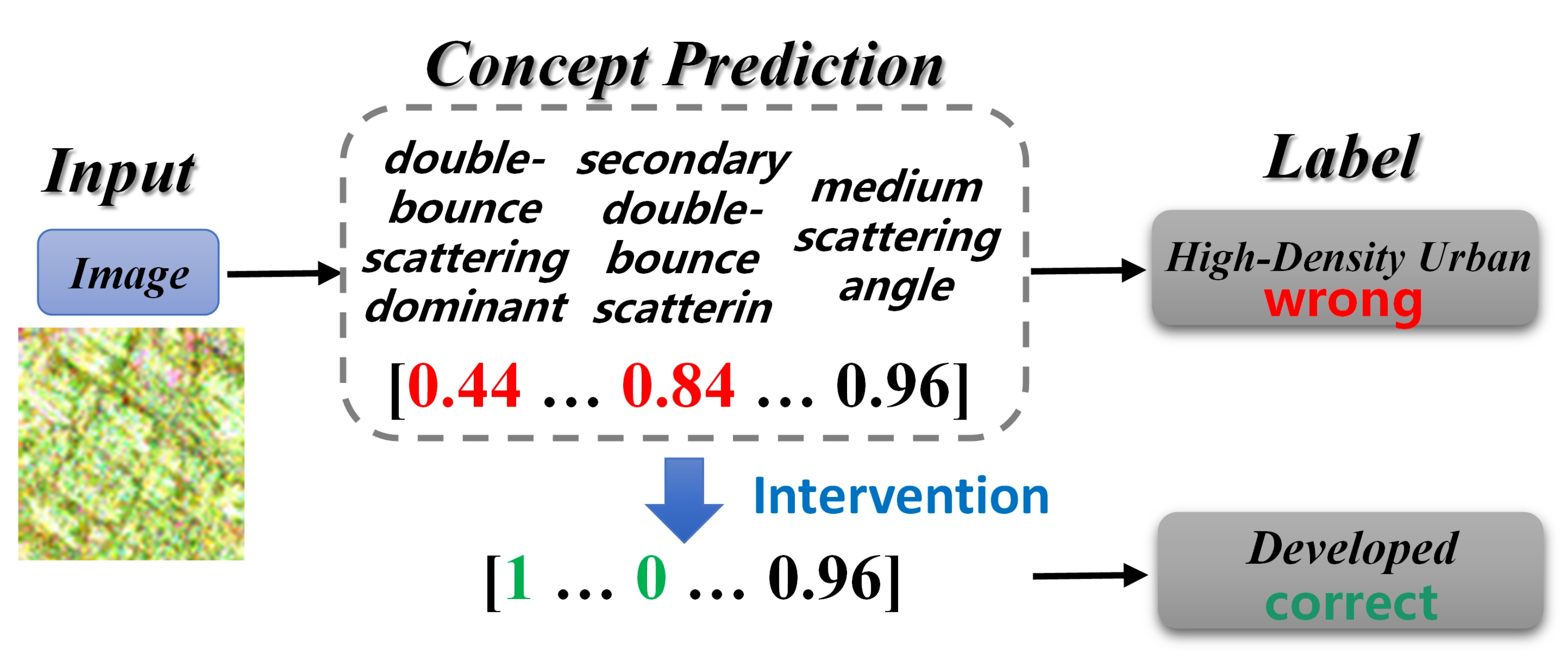}
    \caption{Example of explainable intervention and classification correction}
    \label{fig:intervention}
\end{figure}

Fig. \ref{fig:intervention} illustrates the intervention capability of the CBM in PolSAR image classification. After the raw input image was processed by the model, a set of intermediate semantic concepts—such as “double-bounce scattering dominant,” “secondary double-bounce scattering,” and “medium scattering angle”—were generated at the concept layer and subsequently mapped to a final label. As shown, the model initially misclassified the sample as “High-Density Urban.” By examining the concept layer, we found that certain concept activations (e.g., 0.84) heavily influenced this incorrect prediction. To test the model’s controllability, we intervened by manually adjusting the erroneous concept activations to more semantically accurate values (e.g., changing 0.84 to 0). The revised concept vector became \([1 … 0 … 0.96]\), which was then passed through the concept-to-label mapping module to produce a new prediction. This intervention successfully corrected the classification to “Developed.” These results demonstrate the concept bottleneck layer’s strength in interpretability and controllability. By visualizing and manually modifying intermediate semantic concepts, researchers can precisely identify sources of prediction errors and apply targeted corrections, ultimately enhancing the model’s sensitivity to critical features and improving classification robustness.

\section{conclusion}
In recent years, DL-based algorithms have attracted significant attention in PolSAR image classification due to their remarkable performance. However, the inherent “black-box” nature of DL models, coupled with the limitations of human cognition, poses a major challenge in interpreting the semantic meaning behind the high-dimensional features these models extract. This work aims to improve the interpretability of both the extracted features and the decision-making process in DL-based PolSAR classification. To this end, we propose an effective framework that leverages PTD to reveal the underlying physical scattering mechanisms. Through the carefully designed PaCBM architecture integrated with a KAN, we develop a PolSAR classification model that achieves both strong interpretability and competitive performance. Experimental results on multiple PolSAR datasets demonstrate that our approach not only maintains robust classification accuracy but also enables quantification of the model’s attention to target-relevant polarimetric information. Furthermore, by incorporating spline functions, the model derives category labels from conceptual representations in an interpretable and physically meaningful manner. Although our approach enhances the transparency of DL-based PolSAR classification, it represents a step toward fully “opening the black box.” We hope this work inspires further research on polarization information and interpretability within the PolSAR community.

\bibliographystyle{IEEEtran}
\bibliography{manuscript.bib}

\begin{IEEEbiography}[{\includegraphics[width=1in,height=1.25in,clip,keepaspectratio]{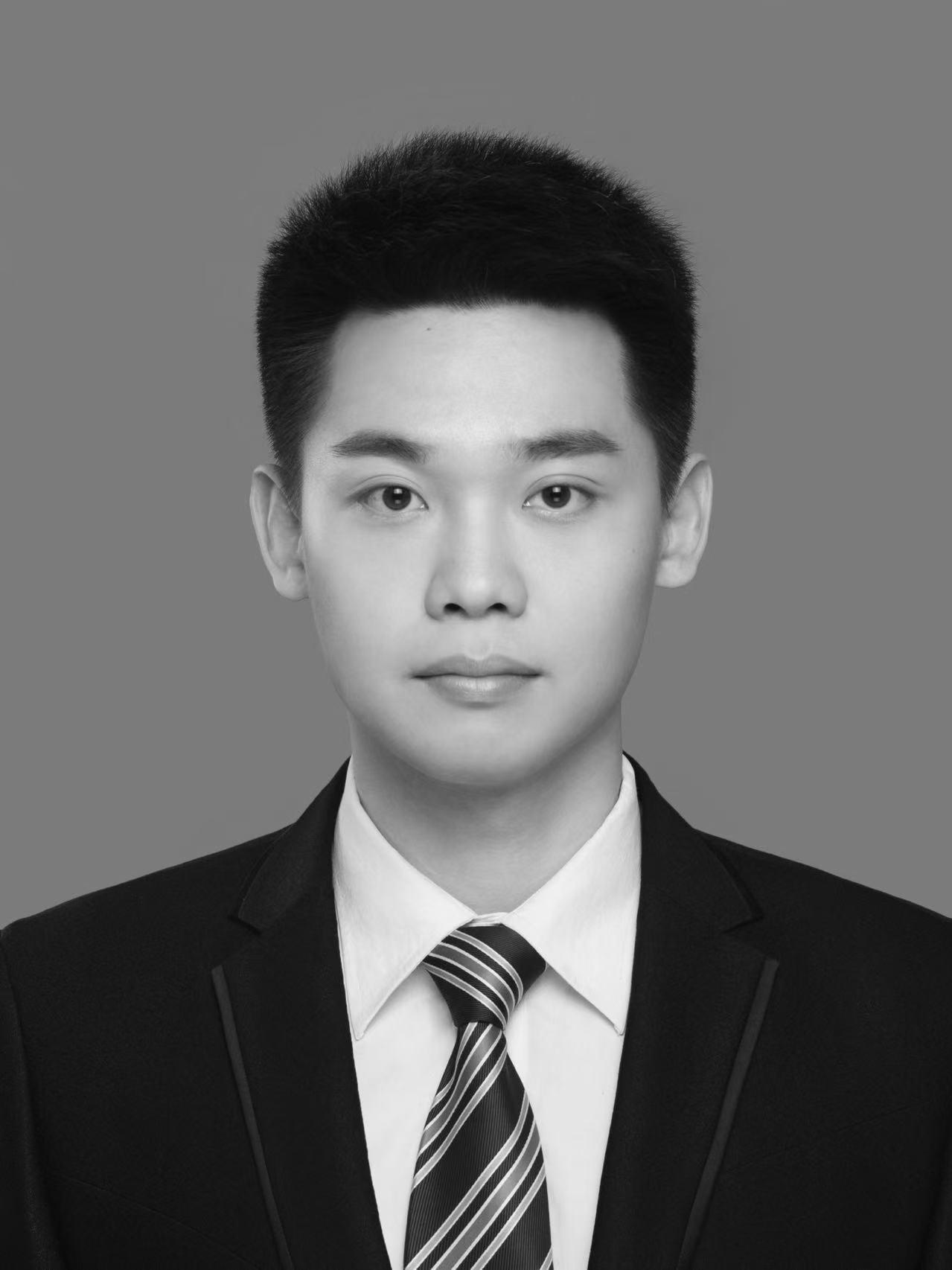}}]{Jinqi Zhang} received the B.E. degree in information and communication engineering from Harbin Institute of Technology, Weihai, China. He is currently pursuing the Ph.D. degree with Harbin Institute of Technology, Harbin, China. 
	
His research interests include deep learning, SAR image processing and interpretability of neural networks.
\end{IEEEbiography}

\begin{IEEEbiography}[{\includegraphics[width=1in,height=1.25in,clip,keepaspectratio]{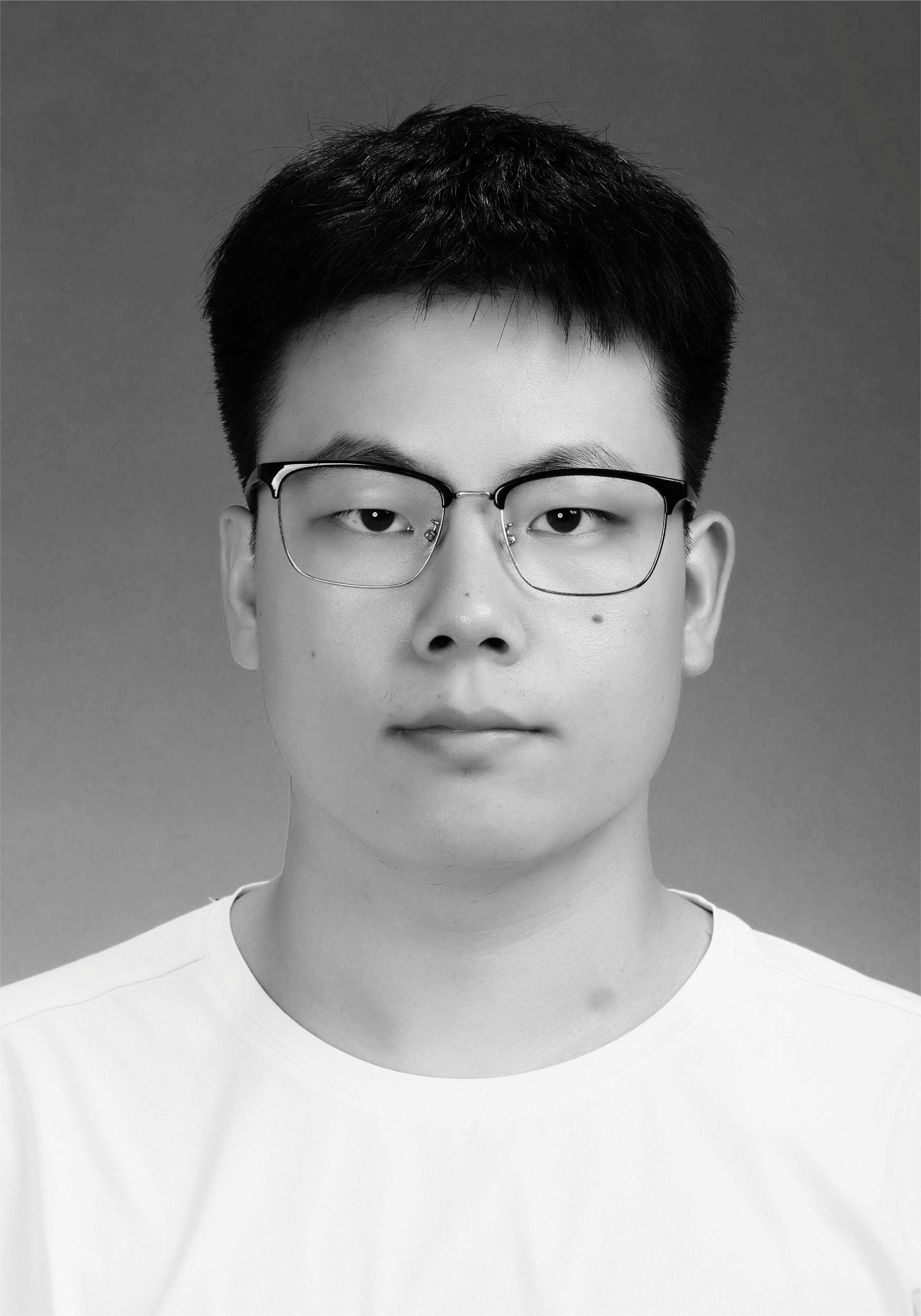}}]{Fangzhou Han} received the B.E. degree in information and communication engineering from Harbin Institute of Technology, Harbin, China. He is currently pursuing the M.S. degree with Harbin Institute of Technology, Harbin, China. 
	
His research interests include deep learning, SAR image processing and causal inference.
\end{IEEEbiography}

\begin{IEEEbiography}[{\includegraphics[width=1in,height=1.25in,clip,keepaspectratio]{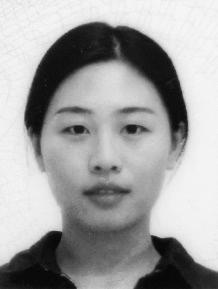}}]{Di Zhuang} was born in Linyi City, Shandong Province, China in 1995. She received the B.S., M.Sc., and Ph.D. degrees in electronic information engineering from the Harbin Institute of Technology, China, in 2018, 2020 and 2025, respectively.

Now, she is now postdoctor in the Department of Information Engineering, School of Electronics and Information Technology, HIT. She currently focuses on SAR image processing, polarimetric data decomposition, InSAR image processing, and PolInSAR technology.
\end{IEEEbiography}

\begin{IEEEbiography}[{\includegraphics[width=1in,height=1.25in,clip,keepaspectratio]{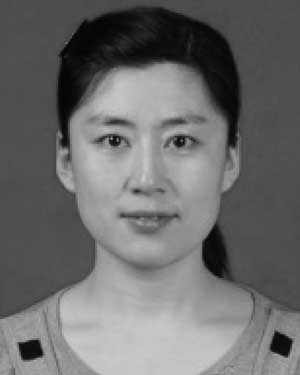}}]{Lamei Zhang} (Senior Member, IEEE) received the B.S., M.Sc., and Ph.D. degrees in information and communication engineering from Harbin Institute of Technology, Harbin, China, in 2004, 2006, and 2010, respectively. 
	
She is currently an Professor with the Department of Information Engineering, Harbin Institute of Technology. Her research interests include remote sensing images processing, information extraction and interpretation of high-resolution synthetic aperture radar, polarimetric SAR, and polarimetric SAR interferometry.

Dr. Zhang serves as the Secretary for the IEEE Harbin Education Section.
\end{IEEEbiography}

\begin{IEEEbiography}[{\includegraphics[width=1in,height=1.25in,clip,keepaspectratio]{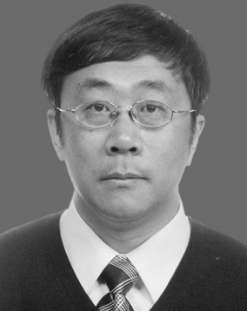}}]{Bin Zou} (Senior Member, IEEE) received the B.S. degree in electronic engineering from the Harbin Institute of Technology, Harbin, China, in 1990, the M.Sc. degree in space studies from the International Space University, Strasbourg, France, in 1998, and the Ph.D. degree in information and communication engineering from the Harbin Institute of Technology, in 2001.
        
From 1990 to 2000, he was with the Department of Space Electrooptic Engineering, Harbin Institute of Technology. From 2003 to 2004, he was a Visiting Scholar with the Department of Geological Sciences, University of Manitoba, Winnipeg, MB, Canada. He is currently a Professor with the Department of Information Engineering, Harbin Institute of Technology. His research interests include SAR image processing, polarimetric SAR, and polarimetric SAR interferometry.
\end{IEEEbiography}

\end{document}